\documentclass{amsart}

\usepackage{amssymb,latexsym}

\numberwithin{equation}{section}

\newtheorem{theorem}{Theorem}[section]
\newtheorem{proposition}[theorem]{Proposition}
\newtheorem{conjecture}[theorem]{Conjecture}
\newtheorem{corollary}[theorem]{Corollary}
\newtheorem{remark}[theorem]{Remark}
\newtheorem{lemma}[theorem]{Lemma}
\newtheorem{example}[theorem]{Example}
\newtheorem{definition}[theorem]{Definition}

\def\proof{\smallskip\noindent {\bf Proof. }}
\def\endproof{\hfill$\square$\medskip}

\def\ZZ{\mathbb{Z}}

\def\RR{\mathbb{R}}
\def\QQ{\mathbb{Q}}
\def\YY{\mathcal{Y}}
\def\ii{\mathbf{i}}

\def\verylongleftrightarrow{\longleftarrow\!\!\!\longrightarrow}
\def\Hugeleftarrow{\!\longleftarrow\!\!\!-\!\!\!-\!\!\!-\!\!\!-\!\!\!-\!\!\!-\!\!\!-\!\!\!-\!\!\!-\!\!\!-\!\!\!-\!\!\!-\!\!\!-\!\!\!-\!\!\!-\!\!\!-}

\begin{document}

\title[$Y$-systems and generalized associahedra]
{$Y$-systems and generalized associahedra}

\author{Sergey Fomin}
\address{Department of Mathematics, University of Michigan,
Ann Arbor, MI 48109, USA} \email{fomin@umich.edu}

\author{Andrei Zelevinsky}
\address{\noindent Department of Mathematics, Northeastern University,
  Boston, MA 02115}
\email{andrei@neu.edu}

\date{November 5, 2001}

\dedicatory{To the memory of Rodica Simion}

\thanks{Research 
supported in part
by NSF grants 
DMS-0070685 (S.F.) and
DMS-9971362 (A.Z.).}


\maketitle




The goals of this paper are two-fold.
First, we prove, for an arbitrary finite root system~$\Phi$, the
periodicity conjecture of Al.~B.~Zamolodchikov~\cite{zamolodchikov}
that concerns \hbox{\emph{$Y$-systems,}} a particular class of
functional relations playing an important role in the theory of
thermodynamic Bethe ansatz.
Algebraically, $Y$-systems can be viewed as families of rational functions
defined by certain birational recurrences formulated in terms of the
root system~$\Phi$.
We obtain explicit formulas for these rational functions,
which turn out to always be Laurent polynomials,
and prove that they
exhibit the periodicity property conjectured
by Zamolodchikov.

In a closely related development, we introduce and
study a simplicial complex~$\Delta(\Phi)$,
which can be viewed as a generalization of the Stasheff
polytope (also known as \emph{associahedron}) for an arbitrary
root system~$\Phi$.
In type~$A$, this complex 
is the face complex of the
ordinary associahedron, whereas in type~$B$,
our construction produces the Bott-Taubes polytope, or cyclohedron.
We enumerate the faces of the complex~$\Delta(\Phi)$,
prove that its geometric realization is always a sphere,
and describe it in  concrete combinatorial terms for the classical
types~$ABCD$.

The primary motivation for this investigation came from the theory of
\emph{cluster algebras,} introduced in~\cite{fz-clust1}
as a device for studying dual canonical bases and total positivity in
semisimple Lie groups.
This connection remains behind the scene in the text of this paper,
and will be brought to light in a forthcoming sequel
to~\cite{fz-clust1}.

\tableofcontents

\section{Main results}

Throughout this paper, $I$ is an $n$-element set of indices, and
$A = (a_{ij})_{i,j \in I}$ an indecomposable Cartan matrix
of finite type; in other words, $A$ is of one of the types $A_n, B_n, \dots, G_2$
on the Cartan-Killing list. Let $\Phi$ be the corresponding root
system (of rank~$n$), and $h$ the Coxeter number.

The first main result of this paper is the following theorem.

\begin{theorem}
\label{th:Bethe-periodicity}
{\rm (Zamolodchikov's conjecture)}
A family $(Y_i (t))_{i \in I, t \in \ZZ}$ of commuting
variables satisfying the recurrence relations
\begin{equation}
\label{eq:y-system}
Y_i(t+1) Y_i(t-1)
= \prod_{j\neq i} (Y_j(t)+1)^{-a_{ij}}
\end{equation}
is periodic with period $2(h+2)$, i.e.,
$Y_i (t + 2(h+2)) = Y_i (t)$ for all $i$ and $t$.
\end{theorem}

We refer to the relations (\ref{eq:y-system}) as the \emph{$Y$-system}
associated with the matrix~$A$ (or with the root system~$\Phi$).
$Y$-systems arise in the theory of thermodynamic Bethe ansatz, as
first shown by Al.~B.~Zamolodchikov~\cite{zamolodchikov}.
The periodicity in Theorem~\ref{th:Bethe-periodicity}
was conjectured  by Zamolodchikov~\cite{zamolodchikov} in the
simply-laced case, i.e., when the product in the right-hand-side of
(\ref{eq:y-system}) is square-free.
The type~$A$ case of Zamolodchikov's conjecture was proved independently
by  E.~Frenkel and A.~Szenes~\cite{frenkel-szenes}
and by F.~Gliozzi and  R.~Tateo~\cite{gliozzi-tateo};
the type~$D$ case was considered in~\cite{CGT}.
This paper does not deal with $Y$-systems more general than (\ref{eq:y-system}),
defined by pairs of Dynkin diagrams (see~\cite{RVT}, \cite{KNS},
and~\cite{kuniba-nakanishi}).

Our proof of Theorem~\ref{th:Bethe-periodicity} is based on the following reformulation.
Recall that the Coxeter graph associated to a Cartan matrix $A$
has the indices in~$I$ as vertices, with $i,j\in I$ joined by an
edge whenever $a_{ij} a_{ji} > 0$.
This graph is a tree, hence is bipartite.
We denote the two parts of $I$ by $I_+$ and $I_-$,
and write $\varepsilon (i) = \varepsilon$ for $i \in I_\varepsilon$.
Let $\QQ(u)$ be the field of rational functions in the variables $u_i \, (i \in I)$.
We introduce the involutive automorphisms $\tau_+$ and
$\tau_-$ of $\QQ(u)$ by setting
\begin{equation}
\label{eq:tau-pm}
\tau_\varepsilon (u_i) =
\begin{cases}
\displaystyle\frac{\prod_{j \neq i} (u_j + 1)^{- a_{ij}}}{u_i} &
\text{if $\varepsilon (i) = \varepsilon$;} \\[.1in]
u_i & \text{otherwise.}
\end{cases}
\end{equation}

\begin{theorem}
\label{th:Y-periodicity}
The automorphism $\tau_-  \tau_+$ of $\QQ(u)$ is of finite order.
More precisely, let $w_\circ$ denote the longest element in
the Weyl group associated to $A$.
Then the order of $\tau_-  \tau_+$ is
equal to $(h+2)/2$ if $w_\circ  = - 1$ , and is equal to $h+2$ otherwise.
\end{theorem}

Theorem~\ref{th:Y-periodicity} is essentially equivalent to
Zamolodchikov's conjecture; here is why.
First, we note that each equation (\ref{eq:y-system}) only involves
the variables $Y_i(k)$ with a fixed ``parity''
$\varepsilon(i)\cdot(-1)^k$.
We may therefore assume, without loss of generality, that our
$Y$-system satisfies the condition
\begin{equation}
\label{eq:convention-Yi}
Y_i(k)=Y_i(k+1) \text{\ \ whenever\ \ } \varepsilon(i)=(-1)^k.
\end{equation}
Let us combine (\ref{eq:y-system}) and (\ref{eq:convention-Yi}) into
\begin{equation}
\label{eq:Y_i(k+1)}
Y_i(k+1)
=
\begin{cases}
\displaystyle\frac{\prod_{j \neq i} (Y_j(k) + 1)^{- a_{ij}}}{Y_i(k)}
& \text{if $\varepsilon (i) = (-1)^{k+1}$;} \\[.1in]
Y_i(k) & \text{if $\varepsilon (i) = (-1)^k$.}
\end{cases}
\end{equation}
Then set $u_i\! =\!Y_i (0)$ for $i\!\in\! I$ and compare
(\ref{eq:tau-pm}) with (\ref{eq:Y_i(k+1)}).
By induction on~$k$, we obtain 
$Y_i (k)\!=\!(\underbrace{\tau_- \tau_+\cdots \tau_\pm}_{\text{$k$ times}})(u_i)$
for all $k\in \ZZ_{\geq 0}$ and $i\in I$, establishing the claim.
(Informally, the map $(\tau_-\tau_+)^m$ can be computed
either by iterations ``from within,'' i.e,
by repeating the substitution of variables
$\tau_-\tau_+\,$, or by iterations ``from the outside,''
via the recursion (\ref{eq:Y_i(k+1)}).)

\begin{example}
\label{example:A2-tau-rational}
Type~$A_2$.
{\rm
Let $\Phi$ be the root system of type~$A_2$, with $I=\{1,2\}$.
Let us set $I_+=\{1\}$ and $I_-=\{2\}$.
Then
\[
\tau_+(u_1)=\frac{u_2+1}{u_1},\quad
\tau_-\tau_+(u_1)=\frac{\frac{u_1+1}{u_2}+1}{u_1}=\frac{u_1+u_2+1}{u_1u_2}\,,
\]
etc. Continuing these calculations, we obtain the following diagram:
\begin{equation}
\label{eq:A2-tau-rational}
\begin{array}{ccc}
u_1 &
\stackrel{\textstyle\tau_+}{\verylongleftrightarrow}
~\displaystyle\frac{u_2+1}{u_1}~
\stackrel{\textstyle\tau_-}{\verylongleftrightarrow}
~\displaystyle\frac{u_1+u_2+1}{u_1u_2}~
\stackrel{\textstyle\tau_+}{\verylongleftrightarrow}
~\displaystyle\frac{u_1+1}{u_2}~
\stackrel{\textstyle\tau_-}{\verylongleftrightarrow}
& u_2 \,.\\[-.03in]
\circlearrowright & & \circlearrowright \\
\tau_- & & \tau_+
\end{array}
\end{equation}
Thus the map $\tau_-\tau_+$ acts by
\begin{equation}
\label{eq:A2-tautau-rational}
\begin{array}{ccc}
u_1 & \longrightarrow ~\displaystyle\frac{u_1+u_2+1}{u_1u_2}~
\longrightarrow & u_2 \\
\uparrow & & \downarrow \\[.05in]
\displaystyle\frac{u_2+1}{u_1} & \Hugeleftarrow & \displaystyle\frac{u_1+1}{u_2}
\end{array}
\end{equation}
and has period $5=h+2$, as prescribed by
Theorem~\ref{th:Y-periodicity}.
To compare, the $Y$-system recurrence (\ref{eq:Y_i(k+1)})
(which incorporates the convention (\ref{eq:convention-Yi}))
has period $10=2(h+2)$:

\smallskip

\begin{center}
\begin{tabular}{c|cccccccc}
{} & $Y_i(0)$ & $Y_i(1)$ & $Y_i(2)$ & $Y_i(3)$ & $Y_i(4)$ & $Y_i(5)$ &$\cdots$
& $Y_i(10)$\\
\hline
\\[-.1in]
$i=1$ & $u_1$ & $\displaystyle\frac{u_2+1}{u_1}$
&$\displaystyle\frac{u_2+1}{u_1}$
& $\displaystyle\frac{u_1+1}{u_2}$& $\displaystyle\frac{u_1+1}{u_2}$&
$u_2$ & $\cdots$ & $u_1$\\[.1in]
\hline
\\[-.1in]
$i=2$ & $u_2$  & $u_2$ & $\displaystyle\frac{u_1+u_2+1}{u_1u_2}$&
$\displaystyle\frac{u_1+u_2+1}{u_1u_2}$ & $u_1$ & $u_1$ & $\cdots$ & $u_2$
\end{tabular}
\end{center}
}
\end{example}

\medskip

\begin{figure}[ht]
\begin{center}
\setlength{\unitlength}{1.5pt}
\begin{picture}(120,80)(-60,-35)
\thicklines

\put(0,0){\circle*{1}}

\put(0,0){\vector(1,0){50}}
\put(0,0){\vector(-1,0){50}}
\put(0,0){\vector(-2,3){25}}
\put(0,0){\vector(2,3){25}}
\put(0,0){\vector(2,-3){25}}

\put(58,0){\makebox(0,0){$\alpha_1$}}
\put(-58,0){\makebox(0,0){$-\alpha_1$}}
\put(38,37.5){\makebox(0,0){$\alpha_1\!+\!\alpha_2$}}
\put(-32,37.5){\makebox(0,0){$\alpha_2$}}
\put(34,-37.5){\makebox(0,0){$-\alpha_2$}}

\end{picture}
\end{center}
\caption{The set $\Phi_{\geq - 1}$  in type $A_2$}
\label{fig:5-roots}
\end{figure}
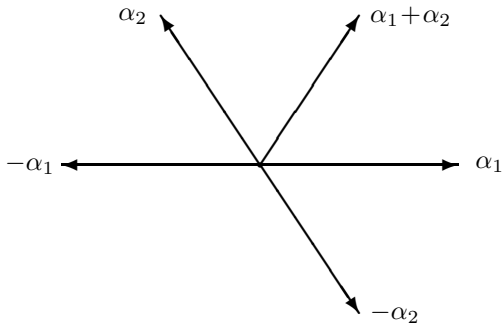

Let $\YY$ denote the smallest set of rational functions that contains
all coordinate functions $u_i$ and is stable under $\tau_+$ and~$\tau_-\,$.
(This set can be viewed as the collection of all distinct variables in a
$Y$-system of the corresponding type.)
For example, in type~$A_2$,
\[
\YY=\left\{u_1, u_2, \displaystyle\frac{u_2+1}{u_1},
\displaystyle\frac{u_1+1}{u_2},
\displaystyle\frac{u_1+u_2+1}{u_1u_2}
 \right\}
\]
(see (\ref{eq:A2-tau-rational})--(\ref{eq:A2-tautau-rational})).
Our proof of Theorem~\ref{th:Y-periodicity} is based on establishing a
bijective correspondence between the set~$\YY$ and a certain
subset $\Phi_{\geq - 1}$ of the root system~$\Phi$;
under this bijection, the involutions $\tau_+$ and $\tau_-$
correspond to some piecewise-linear automorphisms of the ambient
vector space of~$\Phi$, which exhibit the desired periodicity
properties.
To be more precise, let us define
$$\Phi_{\geq - 1} = \Phi_{> 0} \cup (- \Pi) \ ,$$
where $\Pi = \{\alpha_i: i \in I\} \subset \Phi$ is the set of
simple roots, and $\Phi_{> 0}$ the set of positive roots of~$\Phi$.
The case $A_2$ of this definition is illustrated in
Figure~\ref{fig:5-roots}.

\pagebreak[2]

Let $Q=\ZZ\Pi$ be the root lattice, and $Q_\RR$ its ambient real
vector space.
For $\alpha \in Q_\RR$, we denote by $[\alpha : \alpha_i]$
the coefficient of $\alpha_i$ in the expansion of $\alpha$ in the
basis~$\Pi$.
Let $\tau_+$ and $\tau_-$ denote the piecewise-linear
automorphisms of $Q_\RR$ given by
\begin{equation}
\label{eq:tau-pm-tropical}
[\tau_\varepsilon  \alpha: \alpha_i] =
\begin{cases}
- [\alpha: \alpha_i] - \sum_{j \neq i}  a_{ij} \max ([\alpha: \alpha_j], 0)
&
\text{if $\varepsilon (i) = \varepsilon$;} \\[.1in]
[\alpha: \alpha_i] & \text{otherwise.}
\end{cases}
\end{equation}
The reason we use the same symbols for the birational transformations
(\ref{eq:tau-pm}) and the piecewise-linear transformations
(\ref{eq:tau-pm-tropical}) is that the latter can be viewed as the
\emph{tropical specialization} of the former.
This means replacing the usual addition and multiplication
by their tropical versions
\begin{equation}
\label{eq:max-+}
a \,\oplus \,b = {\rm max}\, (a,b) \ , \quad a\, \odot \, b = a + b \ ,
\end{equation}
and replacing the multiplicative unit 1 by~0.

It is easy to show (see Proposition~\ref{pr:tau-t}) that each of the
maps $\tau_\pm$ defined by
(\ref{eq:tau-pm-tropical}) preserves the subset $\Phi_{\geq - 1}$.

\begin{theorem}
\label{th:Y-Phi}
There exists a unique bijection
$\alpha \mapsto Y[\alpha]$ between
$\Phi_{\geq - 1}$ and $\YY$ such that
$Y[- \alpha_i] = u_i$ for all $i \in I$, and
$\tau_\pm (Y[\alpha]) = Y[\tau_\pm (\alpha)]$
for all $\alpha \in \Phi_{\geq - 1}$.
\end{theorem}

Passing from $\YY$ to $\Phi_{\geq -1}$ and from
(\ref{eq:tau-pm}) to (\ref{eq:tau-pm-tropical})
can be viewed as a kind of ``linearization,"
with the important distinction that the action of $\tau_\pm$ in $Q_\RR$
given by (\ref{eq:tau-pm-tropical}) is
piecewise-linear rather than linear.
This ``tropicalization" procedure appeared in some of our previous work
\cite{bfz96, bz01, fz-clust1}, although there it was the birational
version that shed the light on the piecewise-linear one.
In the present context, we go in the opposite direction:
we first prove the tropical version of
Theorem~\ref{th:Y-periodicity} (see Theorem~\ref{th:dihedral}),
and then obtain the original version by combining the tropical one
with Theorem~\ref{th:Y-Phi}.

In the process of proving Theorem~\ref{th:Y-Phi}, we find explicit
expressions for the rational functions $Y[\alpha]$.
It turns out that these functions exhibit the \emph{Laurent phenomenon}
(cf.~\cite{fz-Laurent}), that is, all of them are Laurent polynomials in
the variables~$u_i$.
Furthermore, the denominators of these Laurent polynomials are all
distinct, and are canonically in bijection with the elements of the
set~$\Phi_{\geq -1}$.
More precisely, let $\alpha \mapsto \alpha^\vee$ denote the
natural bijection between $\Phi$ and the dual root system
$\Phi^\vee$, and let us abbreviate
$$u^{\alpha^\vee} = \prod_{i \in I} u_i^{[\alpha^\vee : \alpha^\vee_i]} \ .$$

\begin{theorem}
\label{th:Y-Laurent}
For every root $\alpha \in \Phi_{\geq -1}$, we have
\begin{equation}
\label{eq:Y[alpha]}
Y[\alpha] = \frac{N[\alpha]}{u^{\alpha^\vee}},
\end{equation}
where $N[\alpha]$ is a polynomial in the $u_i$ with positive integer
coefficients and constant term~$1$.
\end{theorem}


To illustrate Theorem~\ref{th:Y-Laurent}: in type~$A_2$,  we have
\[
\begin{array}{lll}
Y[-\alpha_1] = u_1=\displaystyle\frac{1}{u_1^{-1}} & &
Y[\alpha_1] = \displaystyle\frac{u_2+1}{u_1}\\[.2in]
Y[-\alpha_2] = u_2=\displaystyle\frac{1}{u_2^{-1}} & &
Y[\alpha_2] = \displaystyle\frac{u_1+1}{u_2}
\end{array}
\quad
Y[\alpha_1+\alpha_2] =\displaystyle\frac{u_1+u_2+1}{u_1u_2}
\]
In any type, we have
\begin{equation*}
Y[- \alpha_i] = u_i,
\quad
N[- \alpha_i] = 1,
\end{equation*}
\begin{equation*}
Y[\alpha_i] = \tau_{\varepsilon (i)} u_i =
\displaystyle\frac{\prod_{j \neq i} (u_j + 1)^{- a_{ij}}}{u_i} \ ,
\quad
N[\alpha_i] =
\prod_{j \neq i} (u_j + 1)^{- a_{ij}} \ .
\end{equation*}

Each numerator $N[\alpha]$ in (\ref{eq:Y[alpha]}) can be expressed
as a product of ``smaller" polynomials, which are also labeled by
roots from $\Phi_{\geq -1}$.
These polynomials are defined as follows.

\begin{theorem}
\label{th:F-polynomials}
There exists a unique family
$(F[\alpha])_{\alpha \in \Phi_{\geq -1}}$
of polynomials  in the variables $u_i (i\in I)$
such that
\begin{itemize}
\item[{\rm(i)}]
$F[- \alpha_i] = 1$ for all $i \in I$;
\item[{\rm(ii)}]
for any $\alpha \in \Phi_{\geq - 1}$
and any $\varepsilon\in\{+,-\}$, we have
\begin{equation}
\label{eq:tau-F}
\tau_\varepsilon (F[\alpha]) =
\frac{\prod_{\varepsilon (i) = - \varepsilon}
(u_i + 1)^{[\alpha^\vee : \alpha^\vee_i]}}
{\prod_{\varepsilon (i) = \varepsilon}
u_i^{\max ([\alpha^\vee : \alpha^\vee_i],0)}}
\cdot
F[\tau_{- \varepsilon} (\alpha)].
\end{equation}
\end{itemize}
Furthermore, each $F[\alpha]$ is a polynomial in the $u_i$ with
positive integer coefficients and constant term~$1$.
\end{theorem}

We call the polynomials $F[\alpha]$ described in
Theorem~\ref{th:F-polynomials}
the \emph{Fibonacci polynomials} of type~$\Phi$.
The terminology comes from the fact that in the type~$A$ case,
each of these polynomials is a sum of a Fibonacci number
of monomials; cf.\
Example~\ref{example:fib-A}.

In view of Theorem~\ref{th:Y-Phi},
every root $\alpha\in\Phi_{\geq -1}$ can be written as
\begin{equation}
\label{eq:alpha-ki}
\alpha=\alpha(k;i) \stackrel{\rm def}{=} (\tau_-  \tau_+)^k( - \alpha_i)
\end{equation}
for some $k\in\ZZ$ and $i\in I$.

\begin{theorem}
\label{th:N-thru-F}
For $\alpha = \alpha(k;i) \in \Phi_{\geq -1}$, we have
\begin{equation}
\label{eq:N-thru-F}
N[\alpha] = \prod_{j \neq i}
F[ \alpha(-k;j)
]^{- a_{ij}}
 \ .
\end{equation}
\end{theorem}

We conjecture that all polynomials $F[\alpha]$ are irreducible,
so that (\ref{eq:N-thru-F}) provides the irreducible factorization
of~$N[\alpha]$.

Among the theorems stated above, the core result, which implies the
rest (see Section~\ref{sec:Th1.6-implies-Zam}), is
Theorem~\ref{th:F-polynomials}.
This theorem is proved in Section~\ref{sec:fibonacci} according to the
following plan.
We begin by reducing the problem to the simply-laced case by a standard
``folding'' argument.
In the $ADE$ case, the proof is obtained by explicitly writing the
monomial expansions of the polynomials $F[\alpha]$
and checking that the polynomials thus defined satisfy the conditions in
Theorem~\ref{th:F-polynomials}.
This is done in two steps.
First, we give a uniform formula for the monomial expansion of
$F[\alpha]$ whenever
$\alpha=\alpha^\vee$ is a positive root of ``classical type,"
i.e., all the coefficients $[\alpha : \alpha_i]$ are equal to $0$,
$1$, or~$2$ (see (\ref{eq:F-2-restricted})). This in particular
covers the $A$ and $D$ series of root systems. We compute the rest
of the Fibonacci polynomials for the exceptional types $E_6$,
$E_7$, and $E_8$ using \texttt{Maple} (see the last part of
Section~\ref{sec:fibonacci}). In fact, the computational resources
of \texttt{Maple} (on a 16-bit processor) turned out to be barely
sufficient for handling the case of~$E_8$; it seems that for this
type, it would be next to impossible to prove Zamolodchikov's
conjecture by direct calculations based on iterations of the
recurrence (\ref{eq:y-system}).

\medskip

\pagebreak[2]

We next turn to the second group of our results, which concern
a particular simplicial complex~$\Delta (\Phi)$ associated to the root
system~$\Phi$.
This complex has $\Phi_{\geq -1}$ as the set of vertices.
To describe the faces of $\Delta (\Phi)$, we will need the notion of a
\emph{compatibility degree} $(\alpha \| \beta)$ of two roots
$\alpha,\beta\in\Phi_{\geq -1}$.
We define
\begin{equation}
\label{eq:compatibility-degree}
(\alpha \| \beta)=
[Y[\alpha]+1]_{\rm trop} (\beta),
\end{equation}
where $[Y[\alpha]+1]_{\rm trop}$  denotes
the tropical specialization (cf.\ (\ref{eq:max-+}))
of the Laurent polynomial $Y[\alpha]+1$,
which is then evaluated at the $n$-tuple
$(u_i = [\beta:\alpha_i])_{i \in I}$.

We say that two vertices $\alpha$ and $\beta$ are
\emph{compatible} if
$(\alpha \| \beta)=0$.
The compatibility degree can be given a simple alternative definition
(see Proposition~\ref{pr:compatibility-explicit}), which implies,
somewhat surprisingly, that the condition $(\alpha \| \beta)=0$ is
symmetric in $\alpha$ and $\beta$ (see
Proposition~\ref{pr:compatibility-symmetry}).
We then define the simplices of $\Delta (\Phi)$ as mutually compatible
subsets of $\Phi_{\geq -1}$.
The maximal simplices of $\Delta (\Phi)$ are called the \emph{clusters}
associated to~$\Phi$.

To illustrate, in type~$A_2$, the values of $(\alpha \| \beta)$ are
given by the table
\smallskip

\begin{center}
\begin{tabular}{c|ccccc}
 & $-\alpha_1$ & $-\alpha_2$ & \ \ $\alpha_1$\ \  & \ \ $\alpha_2$\ \
 & $\alpha_1\!+\!\alpha_2$ \\
\hline
\\[-.1in]
$-\alpha_1$ & 0 & 0 & 1 & 0 & 1 \\
$-\alpha_2$ & 0 & 0 & 0 & 1 & 1 \\
$\alpha_1$  & 1 & 0 & 0 & 1 & 0 \\
$\alpha_2$  & 0 & 1 & 1 & 0 & 0 \\
$\alpha_1+\alpha_2$ & 1 & 1 & 0 & 0 & 0
\end{tabular}
\end{center}

\medskip

\noindent
The clusters of type $A_2$ are thus given by the list
\[
\{-\alpha_1,\alpha_2\},\
\{\alpha_2,\alpha_1+\alpha_2\},\
\{\alpha_1+\alpha_2, \alpha_1\},\
\{\alpha_1, -\alpha_2\},
\{-\alpha_2,-\alpha_1\}.
\]
Note that these are exactly the pairs of roots represented by adjacent
vectors in Figure~\ref{fig:5-roots}.


\begin{theorem}
\label{th:cluster-purity}
The complex $\Delta (\Phi)$ is pure of
dimension $n\!-\!1$. 
In other words, all clusters are of the same size~$n$.
Moreover, each cluster is a $\ZZ$-basis of the root lattice~$Q$.
\end{theorem}

We obtain recurrence relations for the face numbers of~$\Delta (\Phi)$,
which enumerate simplices of any given dimension (see
Proposition~\ref{pr:f-vector}).
In particular, we compute explicitly the total number of clusters.

\begin{theorem}
\label{th:N-thru-exponents}
For a root system $\Phi$ of a Cartan-Killing type $X_n$, the total
number of clusters is given by the formula
\begin{equation}
\label{eq:N-thru-exponents}
N(X_n) = \prod_{i=1}^n \frac{e_i + h + 1}{e_i + 1} \ ,
\end{equation}
where $e_1, \dots, e_n$ are the exponents of~$\Phi$,
and $h$ is the Coxeter number.
\end{theorem}

Explicit expressions for the numbers $N(X_n)$ for all Cartan-Killing
types~$X_n$ are given in Table~\ref{tab:cluster-numbers}
(Section~\ref{sec:cluster-complexes}).
We are grateful to Fr\'ed\'eric Chapoton who observed that these
expressions, which we obtained on a case by case basis,
can be replaced by the unifying formula (\ref{eq:N-thru-exponents}).
F.~Chapoton also brought to our attention that the numbers in
(\ref{eq:N-thru-exponents}) appear in the study of
non-crossing and non-nesting partitions by V.~Reiner,
C.~Athanasiadis, and A.~Postnikov~\cite{reiner, ath}.
For the classical types $A_n$ and $B_n$, a bijection between
clusters and non-crossing partitions is established in
Section~\ref{sec:clusters-classical}.

We next turn to the geometric realization of $\Delta (\Phi)$.
The reader is referred to \cite{ziegler} for terminology and basic
background on convex polytopes.

\begin{theorem}
\label{th:cluster-fan}
The simplicial cones $\RR_{\geq 0} C$
generated by all clusters $C$ form a complete simplicial fan
in the ambient real vector space~$Q_\RR$:
the interiors of these cones are mutually disjoint, and
the union of these cones is the entire space~$Q_\RR$.
\end{theorem}

\begin{corollary}
\label{cor:sphere}
The geometric realization of the complex $\Delta (\Phi)$ is an
$(n-\!1)$-dimensional sphere.
\end{corollary}

\begin{conjecture}
\label{con:cluster-polytope}
{\rm The simplicial fan in Theorem~\ref{th:cluster-fan}
is the normal fan of a simple $n$-dimensional convex polytope $P(\Phi)$.
}
\end{conjecture}

\begin{figure}[ht]
\begin{center}
\setlength{\unitlength}{2pt}
\begin{picture}(120,85)(-60,-45)
\thicklines

\put(0,0){\circle{20}}

\thinlines

\put(0,0){\circle*{1}}

\put(0,0){\vector(1,0){50}}
\put(0,0){\vector(-1,0){50}}
\put(0,0){\vector(-2,3){25}}
\put(0,0){\vector(2,3){25}}
\put(0,0){\vector(2,-3){25}}

\put(58,0){\makebox(0,0){$\alpha_1$}}
\put(-58,0){\makebox(0,0){$-\alpha_1$}}
\put(38,37.5){\makebox(0,0){$\alpha_1\!+\!\alpha_2$}}
\put(-32,37.5){\makebox(0,0){$\alpha_2$}}
\put(34,-37.5){\makebox(0,0){$-\alpha_2$}}

\thicklines

\put(25,15){\line(0,-1){30}}
\put(25,15){\line(-3,2){25}}
\put(-25,15){\line(3,2){25}}
\put(25,-15){\line(-3,-2){50}}
\put(-25,15){\line(0,-1){63.33}}

\put(10,0){\circle*{2}}
\put(-10,0){\circle*{2}}
\put(5.55,8.32){\circle*{2}}
\put(-5.55,8.32){\circle*{2}}
\put(5.55,-8.32){\circle*{2}}

\put(25,15){\circle*{2}}
\put(25,-15){\circle*{2}}
\put(-25,15){\circle*{2}}
\put(0,31.67){\circle*{2}}
\put(-25,-48.33){\circle*{2}}

\end{picture}
\end{center}
\caption{The complex $\Delta(\Phi)$ and the polytope $P(\Phi)$ in type
  $A_2$}
\label{fig:5-roots-dual}
\end{figure}
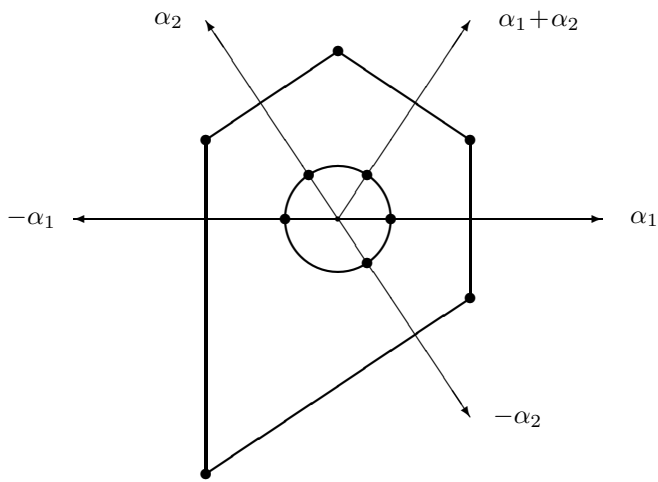

The type $A_2$ case is illustrated in Figure~\ref{fig:5-roots-dual}.

The following is a weaker version of
Conjecture~\ref{con:cluster-polytope}.

\begin{conjecture}
\label{con:cluster-polytope-weak}
{\rm The complex $\Delta (\Phi)$ viewed
as a poset under reverse inclusion is the face lattice of
a simple $n$-dimensional convex polytope $P(\Phi)$.
}
\end{conjecture}

By the Blind-Mani theorem (see, e.g., \cite[Section~3.4]{ziegler}),
the face lattice of a simple polytope $P$ is uniquely determined by
the 1-skeleton (the edge graph) of~$P$.
In our situation, the edge graph $E(\Phi)$ of the (conjectural) polytope
$P(\Phi)$ can be described as follows.

\begin{definition}
\label{def:exchange-graph}
{\rm The \emph{exchange graph} $E(\Phi)$ is an (unoriented) graph
whose vertices are the clusters for the root system~$\Phi$,
with two clusters joined by an edge whenever their
intersection is of cardinality~$n\!-\!1$.}
\end{definition}

The following theorem is a corollary of
Theorem~\ref{th:cluster-fan}.

\begin{theorem}
\label{th:cluster-regularity}
For every cluster $C$ and every element $\alpha \in C$, there is a
unique cluster $C'$ such that $C \cap C' = C - \{\alpha\}$.
Thus, the exchange graph $E(\Phi)$ is regular of degree~$n$:
every vertex in $E(\Phi)$ is incident to precisely $n$ edges.
\end{theorem}

We describe the poset $\Delta (\Phi)$ and the exchange graph $E(\Phi)$
in concrete combinatorial terms for all classical types.
This description in particular implies
Conjecture~\ref{con:cluster-polytope-weak} for the
types $A_n$ and~$B_n$; the posets $\Delta (\Phi)$ and $\Delta (\Phi^\vee)$
are canonically isomorphic, so the statement for the type $C_n$
follows as well.
For the type~$A_n$, the corresponding poset $\Delta (A_n)$ can be identified
with the poset of polygonal subdivisions of a regular convex $(n+3)$-gon
by non-crossing diagonals.
This is known to be the face lattice of the \emph{Stasheff polytope,}
or \emph{associahedron}
(see \cite{stasheff}, \cite{lee}, \cite[Chapter~7]{gkz}).
For the type $B_n$, we identify $\Delta (B_n)$ with the
sublattice of $\Delta (A_{2n-1})$ that consists of
centrally symmetric polygonal subdivisions of a regular convex $2(n+1)$-gon
by non-crossing diagonals.
This is the face lattice of the type~$B$ associahedron introduced by R.~Simion
(see \cite[Section~5.2]{simion} and \cite{simion-B}).
Simion's construction is combinatorially equivalent~\cite{devadoss}
to the ``cyclohedron'' complex of R.~Bott and C.~Taubes~\cite{bott-taubes}.
Polytopal realizations of the cyclohedron were constructed explicitly
by M.~Markl~\cite{markl} and R.~Simion~\cite{simion-B}.

Associahedra of types $A$ and $B$ have a number of remarkable
connections with algebraic geometry \cite{gkz}, topology
\cite{stasheff}, knots and operads \cite{bott-taubes,devadoss},
combinatorics \cite{reiner}, etc.
It would be interesting to extend these connections to the
type~$D$ and  the exceptional types.

\medskip

The primary motivation for this investigation came from the theory of
\emph{cluster algebras,} which we introduced in~\cite{fz-clust1}
as a device for studying dual canonical bases and total positivity in
semisimple Lie groups.
This connection remains behind the scene in the text of this paper,
and will be brought to light in a forthcoming sequel to~\cite{fz-clust1}.

The general layout of the paper is as follows.
The material related to
$Y$-systems is treated in Section~\ref{sec:Y-systems-proofs}; in
particular, Theorems~\ref{th:Y-periodicity}, \ref{th:Y-Phi},
\ref{th:Y-Laurent}, \ref{th:F-polynomials}, and~\ref{th:N-thru-F}
are proved there.
Section~\ref{sec:cluster-complexes} is devoted to the study of
the complexes~$\Delta(\Phi)$, including the proofs of
Theorems~\ref{th:cluster-purity}, \ref{th:N-thru-exponents},
and \ref{th:cluster-fan}.

\bigskip

\textsc{Acknowledgments.}
%
We are grateful to Andr\'as  Szenes for introducing us to
$Y$-systems;
to Alexander Barvinok, Satyan Devadoss, Mikhail Kapranov, Victor
Reiner, John Stembridge, and Roberto Tateo for bibliographical guidance;
and to Fr\'ed\'eric Chapoton for
pointing out the numerological connection between $\Delta(\Phi)$
and non-crossing/non-nesting partitions.

Our work on the complexes $\Delta(\Phi)$ was influenced by
Rodica Simion's beautiful construction~\cite{simion,simion-B}
of the type~$B$ associahedra (see Section~\ref{sec:type-bc}).
We dedicate this paper to Rodica's memory.

\pagebreak[2]

\section{$Y$-systems}
\label{sec:Y-systems-proofs}

\subsection{Root system preliminaries}

We start by laying out the basic terminology and notation related
to root systems, to be used throughout the paper;
some of it has already appeared in the introduction.
In what follows, $A = (a_{ij})_{i,j \in I}$ is
an indecomposable $n\times n$ Cartan matrix of finite type,
i.e., one of the matrices $A_n, B_n, \dots, G_2$ in the
Cartan-Killing classification.
Let $\Phi$ be the corresponding rank $n$
root system with the set of simple roots $\Pi = \{\alpha_i: i \in I\}$.
Let $W$ be the Weyl group of $\Phi$, and
$w_\circ$ the longest element of $W$.

We denote by $\Phi^\vee$ the dual root system with the set of
simple coroots $\Pi^\vee = \{\alpha_i^\vee: i \in I\}$.
The correspondence $\alpha_i \mapsto \alpha_i^\vee$ extends
uniquely to a $W$-equivariant bijection $\alpha \mapsto
\alpha^\vee$ between $\Phi$ and $\Phi^\vee$.
Let $\langle\alpha^\vee,\beta\rangle$ denote the natural pairing
$\Phi^\vee \times \Phi \to \ZZ$.
We adopt the convention
$a_{ij} = \langle\alpha_i^\vee, \alpha_j\rangle$.

Let $Q = \ZZ \Pi$ denote the root lattice,
$Q_+ = \ZZ_{\geq 0} \Pi \subset Q$ the additive semigroup generated
by $\Pi$, and $Q_\RR$ the ambient real vector space.
For every $\alpha \in Q_\RR$, we denote by $[\alpha : \alpha_i]$
the coefficient of $\alpha_i$ in the expansion of $\alpha$ in the
basis of simple roots.
In this notation, the action of simple reflections $s_i \in W$ in $Q$
is given as follows:
\begin{equation}
\label{eq:si-action} [s_i \alpha: \alpha_{i'}] =
\begin{cases}
[\alpha: \alpha_{i'}] & \text{if $i'\neq i$;} \\[.in]
- [\alpha: \alpha_i] - \sum_{j \neq i}  a_{ij} [\alpha: \alpha_j] & \text{if
$i'=i$.}
\end{cases}
\end{equation}

The \emph{Coxeter graph} associated to $\Phi$ has the index set $I$
as the set of vertices, with $i$ and $j$ joined by an
edge whenever $a_{ij} a_{ji} > 0$.
Since we assume that $A$ is indecomposable,
the root system $\Phi$ is irreducible, and the Coxeter graph $I$ is a tree.
Therefore, $I$ is a bipartite graph.
Let $I_+$ and $I_-$ be the two parts of~$I$;
they are determined uniquely up to renaming.
We write $\varepsilon (i) = \varepsilon$ for $i \in I_\varepsilon$.

Let $h$ denote the Coxeter number of $\Phi$, i.e., the order of
any Coxeter element in~$W$.
Recall that a Coxeter element is the product of all simple
reflections $s_i$ (for $i \in I$) taken in an arbitrary order.
Our favorite choice of a Coxeter element $t$
will be the following: take $t = t_- t_+$, where
\begin{equation}
\label{eq:coxeter-t-pm}
t_\pm = \prod_{i \in I_\pm} s_i\,.
\end{equation}
Note that the order of factors in (\ref{eq:coxeter-t-pm}) does not
matter because
$s_i$ and $s_j$ commute whenever $\varepsilon (i) = \varepsilon (j)$.

Let us fix some reduced words $\ii_-$ and $\ii_+$ for the elements
$t_-$ and~$t_+\,$. (Recall that $\ii=(i_1,\dots,i_l)$ is called a
reduced word for $w\in W$ if $w=s_{i_1} \cdots s_{i_l}$ is a
shortest-length factorization of $w$ into simple reflections.)

\begin{lemma}
\label{lem:lusztig-rw}
\cite[Exercise~V.\S6.2]{bourbaki}
The word
\begin{equation}
\label{eq:special-rw}
\ii_\circ\stackrel{\rm def}{=}
\underbrace{\ii_- \ii_+ \ii_- \cdots \ii_\mp
  \ii_\pm}_h
\end{equation}
(concatenation of $h$ segments)
is a reduced word for~$w_\circ\,$.
\end{lemma}

Regarding Lemma~\ref{lem:lusztig-rw}, recall that $h$ is even for all
types except $A_n$ with $n$ even;
in the exceptional case of type $A_{2e}$, we have $h = 2e +1$.

We denote by $\Phi_{> 0}$ the set of positive roots of $\Phi$, and let
$$\Phi_{\geq - 1} = \Phi_{> 0} \cup (- \Pi) \ .$$

\subsection{Piecewise-linear version of a $Y$-system
}

For every $i \in I$, we define
a piecewise-linear modification
$\sigma_i : Q \to Q$ of a simple reflection $s_i$ by setting
\begin{equation}
\label{eq:sigma-i-action}
[\sigma_i \alpha: \alpha_{i'}] =
\begin{cases}
[\alpha: \alpha_{i'}] & \text{if $i'\neq i$;}\\
- [\alpha: \alpha_i]
- \sum_{j \neq i}  a_{ij} \max ([\alpha: \alpha_j], 0)
& \text{if $i'=i$.}
\end{cases}
\end{equation}

\begin{proposition}
\label{pr:sigma-i-properties}
{\ }
\begin{enumerate}
\item
Each map $\sigma_i:Q\to Q$  is an involution.

\item
If $i$ and $j$ are not adjacent in the
Coxeter graph, then $\sigma_i$ and $\sigma_j$ commute with each other.
In particular, this is the case whenever $\varepsilon (i) = \varepsilon (j)$.

\item
Each map $\sigma_i$ preserves the set $\Phi_{\geq -1}$.
\end{enumerate}
\end{proposition}

\proof
Parts 1 and 2 are immediate from the definition.
To prove Part~3, notice that for every $i \in I$ and $\alpha \in \Phi_{\geq -1}$,
we have
\begin{equation}
\label{eq:sigma-Phi}
\hspace{1in}
\sigma_i (\alpha) =
\begin{cases}
\alpha & \text{if $\alpha = - \alpha_j \neq - \alpha_i$;} \\ 
s_i (\alpha) & \text{otherwise.} \hspace{1in} \qed
\end{cases}
\end{equation}

\begin{example}
\label{example:A2-tau-tropical} {\rm In type $A_2$ (cf.\
Example~\ref{example:A2-tau-rational}), the actions of $\sigma_1$
and $\sigma_2$ on $\Phi_{\geq -1}$ are given by
\begin{equation}
\label{eq:A2-tau-tropical}
\begin{array}{ccc}
-\alpha_1 & \stackrel{\textstyle\sigma_1}{\verylongleftrightarrow}
~\alpha_1~ \stackrel{\textstyle\sigma_2}{\verylongleftrightarrow}
~\alpha_1\!+\!\alpha_2~
\stackrel{\textstyle\sigma_1}{\verylongleftrightarrow} ~\alpha_2~
\stackrel{\textstyle\sigma_2}{\verylongleftrightarrow}
& -\alpha_2\,. \\
\circlearrowright & & \circlearrowright \\ \sigma_2 & & \sigma_1
\end{array}
\end{equation}
}
\end{example}

By analogy with (\ref{eq:coxeter-t-pm}), we introduce the
piecewise-linear transformations $\tau_+$ and $\tau_-$ of $Q$
by setting
\begin{equation}
\label{eq:coxeter-tau-pm}
\tau_\pm = \prod_{i \in I_\pm} \sigma_i\,;
\end{equation}
this is well-defined in view of
Proposition~\ref{pr:sigma-i-properties}.2.
This definition is of course equivalent to
(\ref{eq:tau-pm-tropical}).
The following properties are easily checked.

\begin{proposition}
\label{pr:tau-t}
{\ }
\begin{enumerate}
\item
Both transformations $\tau_+$ and $\tau_-$ are involutions
and preserve $\Phi_{\geq -1}$.

\item
We have $\tau_\pm (\alpha) = t_\pm (\alpha)$
for any $\alpha \in Q_+$.

\item
The bijection $\alpha \mapsto \alpha^\vee$ between
$\Phi_{\geq -1}$ and $\Phi^\vee_{\geq -1}$ is
$\tau_\pm$-equivariant.
\end{enumerate}
\end{proposition}

It would be interesting to study the group of
piecewise-linear transformations of $Q_\RR$
generated by all the~$\sigma_i$.
In this paper, we focus our attention on the
subgroup of this group generated by the involutions
$\tau_-$ and $\tau_+$.

For $k\in\ZZ$ and $i\in I$, we abbreviate
\[
\alpha(k;i) = (\tau_-  \tau_+)^k( - \alpha_i)
\]
(cf.\ (\ref{eq:alpha-ki})).
In particular, $\alpha(0;i)=-\alpha_i$ for all~$i$
and $\alpha(\pm 1;i)=\alpha_i$ for $i\in I_\mp\,$.

Let $i \mapsto i^*$ denote
the involution on $I$ defined by $w_\circ (\alpha_i) = - \alpha_{i^*}$.
It is known that this involution preserves each of the sets $I_+$ and $I_-$
when $h$ is even, and interchanges them when $h$ is odd.

\medskip
\pagebreak[2]

\begin{proposition}
\label{pr:tau orbits}
{\ }
\begin{enumerate}
\item
Suppose $h = 2e$ is even.
Then the map $(k,i) \mapsto \alpha(k;i)$
restricts to a bijection
\[
[0,e] \times I \to \Phi_{\geq -1} \ .
\]
Furthermore, $\alpha(e+1;i) = - \alpha_{i^*}$
for any~$i$.

\item
Suppose $h = 2e+1$ is odd.
Then the map $(k,i) \mapsto \alpha(k;i)$
restricts to a bijection
\[
([0,e+1] \times I_-) \textstyle\bigcup\, ([0,e] \times I_+)\to
\Phi_{\geq -1} \ .
\]
Furthermore, $\alpha(e+2;i) = - \alpha_{i^*}$
for $i \in I_-$,
and $\alpha(e+1;i) = - \alpha_{i^*}$
for~$i \in I_+$.
\end{enumerate}
\end{proposition}

To illustrate Part 2 of Proposition~\ref{pr:tau orbits},
consider type~$A_2$
(cf.\ (\ref{eq:A2-tau-tropical})).
Then $\tau_-\tau_+=\sigma_2 \sigma_1$ acts on
$\Phi_{\geq -1}$~by
\begin{equation}
\label{eq:A2-tautau-tripical}
\begin{array}{ccc}
-\alpha_1 & \longrightarrow \quad\alpha_1+\alpha_2\quad
\longrightarrow & -\alpha_2 \\[.05in]
\uparrow & & \downarrow \\[.05in]
\alpha_1 & \Hugeleftarrow & \ \alpha_2 \,.
\end{array}
\end{equation}
We thus have
\begin{equation}
\label{eq:A2-wheels}
\begin{array}{lcl}
\alpha(0;1)=-\alpha_1 && \alpha(0;2)=-\alpha_2 \\[.1in]
\alpha(1;1)=\alpha_1+\alpha_2 && \alpha(1;2)=\alpha_2 \\[.1in]
\alpha(2;1)=-\alpha_2 && \alpha(2;2)=\alpha_1 \\[.1in]
\alpha(3;1)=\alpha_2 && \alpha(3;2)=-\alpha_1 .
\end{array}
\end{equation}
To illustrate Part 1 of Proposition~\ref{pr:tau orbits}:
in type~$A_3$, with the standard numbering of roots, we have
\begin{equation}
\label{eq:A3-wheels}
\begin{array}{lclcl}
\alpha(0;1)=-\alpha_1 && \alpha(0;2)=-\alpha_2 && \alpha(0;3)=-\alpha_3 \\[.1in]
\alpha(1;1)=\alpha_1+\alpha_2 && \alpha(1;2)=\alpha_2 && \alpha(1;3)=\alpha_2+\alpha_3 \\[.1in]
\alpha(2;1)=\alpha_3 && \alpha(2;2)=\alpha_1+\alpha_2+\alpha_3 && \alpha(2;3)=\alpha_1 \\[.1in]
\alpha(3;1)=-\alpha_3 && \alpha(3;2)=-\alpha_2 &&
\alpha(3;3)=-\alpha_1.
\end{array}
\end{equation}

\proof
We shall use the following well known fact: for every
reduced word $\ii = (i_1, \dots, i_m)$ of $w_\circ\,$,
the sequence of roots
$\alpha^{(k)} = s_{i_{1}} \cdots s_{i_{k-1}} \alpha_{i_k}$,
for $k = 1, 2, \dots, m$,
is a permutation of~$\Phi_{> 0}\,$ (in particular,
$m=|\Phi_{> 0}|$).
Let $\ii=\ii_\circ$ be the reduced word defined in (\ref{eq:special-rw}).
Direct check shows that in the case when $h = 2e$ is even,
the corresponding sequence of positive
roots $\alpha^{(k)}$
has the form
$$\alpha(1;1), \dots, \alpha(1;n), \alpha(2;1),
\dots, \alpha(2;n), \dots, \alpha(e;1),
\dots, \alpha(e;n) .$$
This implies the first statement in Part~1, and also shows that
$$ \alpha(e;i)=
\begin{cases}
t_- (t_+ t_-)^{e-1} (\alpha_i) & \text{if $i \in I_+$;}\\
(t_- t_+)^{e-1} (\alpha_i) & \text{if $i \in I_-$}
\end{cases}
$$
(recall that $t_-$ and $t_+$ are defined by (\ref{eq:coxeter-t-pm})).
Then, for~$i \in I_+$, we have
\begin{eqnarray*}
\alpha(e+1;i)&=&\tau_-\tau_+ t_- (t_+ t_-)^{e-1} (\alpha_i)
= \tau_-\tau_+ t_- (t_+ t_-)^{e-1} w_\circ (- \alpha_{i^*})\\
&=&  \tau_-\tau_+ t_+ (- \alpha_{i^*})
= -\alpha_{i^*}\,,
\end{eqnarray*}
whereas for $i \in I_-$, we have
\begin{eqnarray*}
\alpha(e+1;i)&=&
\tau_-\tau_+ (t_- t_+)^{e-1} (\alpha_i)
= \tau_-\tau_+ (t_- t_+)^{e-1} w_\circ (- \alpha_{i^*})\\
&=&  \tau_-\tau_+ t_+ t_-(- \alpha_{i^*})
= -\alpha_{i^*} \,.
\end{eqnarray*}
This proves the second statement in Part~1.

The proof of Part~2 is similar.
\endproof

As an immediate corollary of Proposition~\ref{pr:tau orbits},
we obtain the following tropical version of Theorem~\ref{th:Y-periodicity}.
Let $D$ denote the
group of permutations of $\Phi_{\geq -1}$ generated by $\tau_-$ and $\tau_+$.

\begin{theorem}
\label{th:dihedral}
{\ }
\begin{enumerate}
\item
Every $D$-orbit in  $\Phi_{\geq - 1}$ has a
nonempty intersection with $- \Pi$.
More specifically, the correspondence
$\Omega \mapsto \Omega \cap (- \Pi)$ is a bijection between
the $D$-orbits in  $\Phi_{\geq - 1}$ and the $\langle - w_\circ \rangle$-orbits
in~$(- \Pi)$.

\item
The order of $\tau_-  \tau_+$ in $D$ is
equal to $(h+2)/2$ if $w_\circ = -1$, and is equal to $h+2$ otherwise.
Accordingly, $D$ is the dihedral group of order $(h+2)$ or $2(h+2)$.
\end{enumerate}
\end{theorem}

To illustrate, consider the case of type~$A_2$ (cf.\
(\ref{eq:A2-tau-tropical}), (\ref{eq:A2-tautau-tripical})).
Then $D$ is the dihedral group of order~10, given by
\[
D=\langle \tau_+,\tau_- :
\tau_-^2=\tau_+^2=(\tau_-\tau_+)^5=1\rangle\,.
\]

\subsection{Theorem~\ref{th:F-polynomials} implies Zamolodchikov's conjecture}
\label{sec:Th1.6-implies-Zam}

In this section, we show that Theorem~\ref{th:F-polynomials}
implies Theorems~\ref{th:Bethe-periodicity},
\ref{th:Y-periodicity}, \ref{th:Y-Phi},
\ref{th:Y-Laurent}, and~\ref{th:N-thru-F}.
Thus, we assume the existence of a family of \emph{Fibonacci polynomials}
$(F[\alpha])_{\alpha \in \Phi_{\geq -1}}$
in the variables $u_i (i\in I)$ satisfying the conditions in
Theorem~\ref{th:F-polynomials}.

As explained in the introduction,
Theorem~\ref{th:Bethe-periodicity} is a corollary of
Theorem~\ref{th:Y-periodicity}.
In turn, Theorem~\ref{th:Y-periodicity} is obtained by combining
Theorem~\ref{th:Y-Phi} with Theorem~\ref{th:dihedral}.

As for Theorems~\ref{th:Y-Phi}, \ref{th:Y-Laurent} and
\ref{th:N-thru-F}, we are going to obtain them simultaneously,
as parts of a single package.
Namely, we will define the polynomials $N[\alpha]$
by (\ref{eq:N-thru-F}), then define the Laurent polynomials
$Y[\alpha]$ by (\ref{eq:Y[alpha]}), and then show that these
$Y[\alpha]$ satisfy the conditions in Theorems~\ref{th:Y-Phi}.

Our first task is to prove that the correspondence $\alpha \mapsto N[\alpha]$
is well defined, i.e., the right-hand side of
(\ref{eq:N-thru-F}) depends only on $\alpha$, not on the
particular choice of $k$ and $i$ such that $\alpha = \alpha(k;i)$.
To this end, for every $k \in \ZZ$ and $i \in I$, let us denote
$$\Psi(k;i) = \{(\alpha(-k;j), - a_{ij}): j \in I - \{i\},
a_{ij} \neq 0\} \subset \Phi_{\geq -1} \times \ZZ_{>0} \ .
$$
This definition is given with (\ref{eq:N-thru-F}) in view:
note that the latter can be restated as
\begin{equation}
\label{eq:N-thru-F-restated}
N[\alpha(k;i)]=\prod_{(\beta, d) \in \Psi(k;i)}
F[\beta]^d .
\end{equation}

\begin{lemma}
\label{lem:N-thru-F-rhs}
{\ }
\begin{enumerate}
\item
The set $\Psi(k;i)$ depends only on the root $\alpha =
\alpha(k;i)$, hence it can and will be denoted by $\Psi (\alpha)$.

\item
For every $\alpha \in \Phi_{\geq -1}$ and every sign
$\varepsilon$,
we have
\begin{equation}
\label{eq:tau-Psi}
\Psi(\tau_{\varepsilon} \alpha) = \{(\tau_{-\varepsilon} \beta, d):
(\beta, d) \in \Psi(\alpha)\} \ .
\end{equation}

\item
For every $\alpha \in \Phi_{\geq -1}$, we have
\begin{equation}
\label{eq:sum-Psi}
\sum_{(\beta, d) \in \Psi(\alpha)} d \beta^\vee =
t_- \alpha^\vee + t_+ \alpha^\vee \ .
\end{equation}
\end{enumerate}
\end{lemma}

\proof
Parts 1 and 2 follow by routine inspection from Proposition~\ref{pr:tau orbits}.
To prove Part~3, we first check that it holds for $\alpha = \mp \alpha_i$
for some $i \in I$.
Indeed, we have
$$\Psi (\mp \alpha_i) = \{(\mp \alpha_j, - a_{ij}) : j \in I - \{i\},
a_{ij} \neq 0\} \ .$$
Therefore
$$t_- (\mp \alpha_i^\vee) + t_+ (\mp \alpha_i^\vee) =
\pm \alpha_i^\vee \mp (\alpha_i^\vee - \sum_{j \neq i} a_{ij}
\alpha_j^\vee) = \sum_{(\beta, d) \in \Psi(\mp \alpha_i)} d \beta^\vee \ ,$$
as claimed.
It remains to show that if (\ref{eq:sum-Psi}) holds for some
positive root $\alpha$, then it also holds for $\tau_\pm \alpha$.
To see this, we notice that, by Proposition~\ref{pr:tau-t}.2, we have
$t_\pm \beta^\vee = \tau_\pm \beta^\vee$ for $\beta \in \Phi_{\geq 0}$.
Using (\ref{eq:tau-Psi}), we then obtain
$$\sum_{(\beta, d) \in \Psi(\tau_{\varepsilon} \alpha)} d \beta^\vee =
t_{-\varepsilon} \sum_{(\beta, d) \in \Psi(\alpha)} d \beta^\vee =
\alpha^\vee + t_{-\varepsilon} t_{\varepsilon} \alpha^\vee =
(t_- + t_+)  \tau_{\varepsilon} \alpha^\vee \ ,$$
as desired.
\endproof

By Lemma~\ref{lem:N-thru-F-rhs}.1, the polynomials $N[\alpha]$
are well defined by the formula
\[
N[\alpha]=\prod_{(\beta, d) \in \Psi(\alpha)} F[\beta]^d
\]
(cf.\ (\ref{eq:N-thru-F-restated})),
for every $\alpha \in \Phi_{\geq -1}$.
We then set
\begin{equation}
\label{eq:Y-factorization}
Y[\alpha] = \frac{\prod_{(\beta, d) \in \Psi(\alpha)}
F[\beta]^d}{u^{\alpha^\vee}} \ .
\end{equation}
In particular, $Y[- \alpha_i] = u_i$ for all $i$.
Since all the Laurent polynomials $Y[\alpha]$ defined by
(\ref{eq:Y-factorization}) have different denominators, we
conclude that the correspondence
$\alpha \mapsto Y[\alpha]$ is injective.
To complete the proof of Theorems~\ref{th:Y-Phi}, \ref{th:Y-Laurent} and
\ref{th:N-thru-F}, it remains to verify
the relation
$\tau_\pm (Y[\alpha]) = Y[\tau_\pm (\alpha)]$ for $\alpha \in \Phi_{\geq -1}$.

For any sign $\varepsilon$, we introduce the notation
\[
C_\varepsilon (\beta) = \frac{\prod_{j\in I_{-\varepsilon}}
(u_j + 1)^{[\beta^\vee : \alpha^\vee_j]}}
{\prod_{i\in I_\varepsilon}
u_i^{\max ([\beta^\vee : \alpha^\vee_i],0)}}
\]
and use it to rewrite (\ref{eq:tau-F}) as
\[
\tau_\varepsilon(F[\alpha])=C_\varepsilon(\alpha)
F[\tau_{-\varepsilon}(\alpha)]\,.
\]
Together with (\ref{eq:Y-factorization}) and (\ref{eq:tau-Psi}), this
implies
\[
\tau_\varepsilon(Y[\alpha])
= \displaystyle\frac{\prod_{(\beta, d) \in \Psi(\alpha)}
  C_\varepsilon(\beta)^d
  F[\tau_{-\varepsilon}(\beta)]^d}{\tau_\varepsilon(u^{\alpha^\vee})}
= \frac{N[\tau_\varepsilon \alpha]}{\tau_\varepsilon(u^{\alpha^\vee})}
\prod_{(\beta, d) \in \Psi(\alpha)}
  C_\varepsilon(\beta)^d .
\]
Thus, it remains to verify the identity
\begin{equation}
\label{eq:equating-coefficients}
\prod_{(\beta, d) \in \Psi(\alpha)} C_\varepsilon (\beta)^d =
\frac{\tau_\varepsilon (u^{\alpha^\vee})}{u^{\tau_\varepsilon \alpha^\vee}} \ .
\end{equation}

Using (\ref{eq:tau-pm}), (\ref{eq:tau-pm-tropical}),
(\ref{eq:si-action}), and (\ref{eq:coxeter-t-pm}), we calculate the
right-hand side of (\ref{eq:equating-coefficients}) as follows:
\begin{eqnarray}
\nonumber
\frac{\tau_\varepsilon (u^{\alpha^\vee})}{u^{\tau_\varepsilon
    \alpha^\vee}}
&=&
\displaystyle\frac{\prod_{i\in I_{-\varepsilon}}
    u_i^{[\alpha^\vee:\alpha_i^\vee]}
  \prod_{j\in I_{-\varepsilon}}
    (u_j+1)^{- \sum_{i \neq j} a_{ij} [\alpha^\vee : \alpha^\vee_i]}}
{\prod_{i\in I}
    u_i^{[\tau_\varepsilon \alpha^\vee:\alpha_i^\vee]}
  \prod_{i\in I_{\varepsilon}}
    u_i^{[\alpha^\vee:\alpha_i^\vee]}}\\
\label{eq:rhs-2.15}
&=&
\displaystyle\frac{\prod_{j\in I_{-\varepsilon}}
    (u_j+1)^{[t_-\alpha^\vee+t_+\alpha^\vee:\alpha_j^\vee]}}
{\prod_{i\in I_{\varepsilon}}
    u_i^{[\alpha^\vee+\tau_\varepsilon \alpha^\vee:\alpha_i^\vee]}}\,.
\end{eqnarray}
On the other hand, the left-hand side of
(\ref{eq:equating-coefficients}) is given by
\begin{equation}
\label{eq:lhs-2.15}
\prod_{(\beta, d) \in \Psi(\alpha)} C_\varepsilon (\beta)^d
=
\displaystyle\frac{\prod_{j\in I_{-\varepsilon}}
    (u_j+1)^{\sum_{(\beta, d) \in \Psi(\alpha)} d \, [\beta^\vee:\alpha_j^\vee]}}
{\prod_{i\in I_{\varepsilon}}
  u_i^{\sum_{(\beta, d) \in \Psi(\alpha)} d \max([\beta^\vee:\alpha_i^\vee],0)}
}\,.
\end{equation}
The expressions (\ref{eq:rhs-2.15}) and (\ref{eq:lhs-2.15}) are indeed
equal, for the following reasons.
Their numerators are equal by virtue of (\ref{eq:sum-Psi}).
If $\alpha$ is a positive root, then the equality of denominators
follows again from (\ref{eq:sum-Psi})
(note that all the roots $\beta$ are positive as well), whereas
if $\alpha \in - \Pi$, then both denominators are equal to~$1$.

This completes the derivation of Theorems~\ref{th:Y-Phi}, \ref{th:Y-Laurent} and
\ref{th:N-thru-F} (which in turn imply
Theorems~\ref{th:Bethe-periodicity} and~\ref{th:Y-periodicity})
 from Theorem~\ref{th:F-polynomials}.

\begin{remark}
{\rm
The Laurent polynomial $Y[\alpha] + 1$ has a
factorization similar to the factorization of $Y[\alpha]$ given by
(\ref{eq:Y-factorization}):
\begin{equation}
\label{eq:Y+1-thru-F}
Y[\alpha] + 1 =
\frac{F[\tau_+ \alpha]F[\tau_- \alpha]}
{\prod_{i \in I} u_i^{\max ([\alpha^\vee: \alpha_i^\vee],0)}} \ .
\end{equation}
This can be deduced from Theorem~\ref{th:F-polynomials} by an
argument similar to the one given above.
}
\end{remark}

\subsection{Fibonacci polynomials}
\label{sec:fibonacci}
In this section we prove Theorem~\ref{th:F-polynomials}.
We proceed in three steps.

\medskip

\noindent {\bf Step 1. Reduction to the simply-laced case.}
This is done by means of the well known \emph{folding}
procedure---cf., e.g., \cite[1.87]{FSS}, although we use a different
convention (see (\ref{eq:folding}) below).
Let $\tilde \Phi$ be a simply laced irreducible root system
(i.e., one of type $A_n, D_n, E_6, E_7$, or~$E_8$)
with the index set $\tilde I$,
the set of simple roots $\tilde \Pi$, etc.
Suppose $\rho$ is an automorphism  of the Coxeter graph
$\tilde I$ that preserves the parts $\tilde I_+$ and~$\tilde I_-\,$.
Let $I = \tilde I/\langle \rho \rangle$ be the set of
$\rho$-orbits in $\tilde I$, and let $\pi: \tilde I \to I$ be the
canonical projection.
We denote by the same symbol $\pi$ the projection of polynomial rings
\begin{equation}
\label{eq:pi-polynomial}
\begin{array}{rcl}
\ZZ [u_{\tilde i}: \tilde i \in \tilde I] &\longrightarrow&
\ZZ [u_{i}: i \in I]\\[.1in]
u_{\tilde i} &\longmapsto& u_{\pi (\tilde i)}.
\end{array}
\end{equation}
The ``folded" Cartan matrix $A = (a_{ij})_{i,j \in I}$ is defined as follows:
for $i \in I$, pick some $\tilde i \in \tilde I$
such that $\pi (\tilde i) = i$, and set
$(-a_{ij})$ for $j \neq i$ to be the number of indices
$\tilde j \in \tilde I$ such that $\pi (\tilde j) = j$, and
$\tilde j$ is adjacent to $\tilde i$ in~$\tilde I$.
It is known (and easy to check) that $A$ is of finite type, and
that all non-simply-laced indecomposable Cartan matrices can be
obtained this way:
\begin{equation}
\label{eq:folding}
A_{2n-1} \to B_n\,,\quad
D_{n+1} \to C_n\,,\quad
E_6 \to F_4\,,\quad
D_4 \to G_2 \,.
\end{equation}
The mapping $\tilde \Pi^\vee \to \Pi^\vee$ sending
each $\alpha_{\tilde i}^\vee$ to $\alpha_{\pi (\tilde i)}^\vee$
extends by linearity to a surjection $\tilde \Phi^\vee \to \Phi^\vee$,
which we will also denote by $\pi$.
With even more abuse of notation, we also denote by $\pi$
the surjection $\tilde \Phi \to \Phi$
such that $(\pi (\tilde \alpha))^\vee = \pi (\tilde \alpha^\vee)$.
Note that $\rho$ extends naturally to an automorphism of the root system
$\tilde \Phi$, and the fibers of the projection $\pi: \tilde \Phi \to \Phi$
are the $\rho$-orbits on $\tilde \Phi$.
Also, $\pi$ restricts to a surjection
$\tilde \Phi_{\geq -1} \to \Phi_{\geq -1}$,
and we have $\pi \circ \tilde \tau_\pm = \tau_\pm \circ \pi$.

The following proposition follows at once from the above
description.

\begin{proposition}
\label{pr:folding}
Suppose that a family of polynomials
$(F[\tilde \alpha])_{\tilde \alpha \in \tilde \Phi_{\geq -1}}$
in the variables $u_{\tilde i} \, (\tilde i \in \tilde I)$
satisfies the conditions in Theorem~\ref{th:F-polynomials} for
a simply laced root system~$\tilde \Phi$.
Let $\Phi$ be the ``folding'' of $\tilde\Phi$, as described above.
Then the polynomials $(F[\alpha])_{\alpha \in \Phi_{\geq -1}}$
in the variables $u_i \, (i \in I)$ given by
$F[\alpha] = \pi (F[\tilde \alpha])$
(cf.~{\rm (\ref{eq:pi-polynomial})}), where
$\tilde \alpha \in \tilde \Phi_{\geq -1}$ is any root such that
$\pi (\tilde \alpha) = \alpha$, are well-defined,
and satisfy the conditions in Theorem~\ref{th:F-polynomials}.
\end{proposition}

Thus, it is enough to calculate the Fibonacci polynomials of types
$ADE$, and verify that they have the desired properties.
For the other types, these polynomials can be obtained by simply
identifying the variables $u_{\tilde i}$ which fold into the same
variable~$u_i\,$.

\medskip

\noindent {\bf Step 2. Types $A$ and $D$.}
We will now give an explicit formula for the Fibonacci polynomials
$F[\alpha]$ in the case when $\Phi$ is the root system of type
$A_n$ or~$D_n$.
Recall that these systems have the property that
$[\alpha: \alpha_i] \leq 2$ for every $\alpha \in \Phi_{> 0}$ and
every $i \in I$.
Let us fix a positive root $\alpha$ and abbreviate $a_i = [\alpha: \alpha_i]$.
We call a vector $\gamma = \sum_i c_i \alpha_i$ of the root
lattice \emph{$\alpha$-acceptable} if it satisfies the following three conditions:
\begin{enumerate}
\item[(1)]
$0 \leq c_i \leq a_i$ for all $i$;

\item[(2)]
$c_i+c_j\leq 2$ for any adjacent vertices $i$ and~$j$;

\item[(3)]
there is no simple path (ordinary or closed) $(i_0, i_1, \cdots,
i_m)$, $m\geq 1$,
with $c_0=c_1=\dots=c_m=1$ and $a_0=a_m=1$.
\end{enumerate}
In condition~3 above, by a simple path we mean any path in the Coxeter graph whose all
vertices are distinct, except that we allow for~$i_0=i_m\,$.

As before, we abbreviate $u^\gamma = \prod_i u_i^{c_i}$.

\begin{proposition}
\label{pr:F-multiplicity-1-2}
Theorem~\ref{th:F-polynomials} holds
when $\Phi$ is of the type $A_n$ or $D_n$.
In this case, for every
positive root $\alpha = \sum a_i \alpha_i$, we have
\begin{equation}
\label{eq:F-2-restricted}
F[\alpha] = \sum_\gamma 2^{e(\gamma;\alpha)} u^\gamma \ ,
\end{equation}
where the sum is over all $\alpha$-acceptable $\gamma \in Q$,
and $e(\gamma;\alpha)$ is the number
of connected components of the set $\{i \in I: c_i\! =\! 1\}$ that are
contained in $\{i \in I: a_i \!=\! 2\}$.
\end{proposition}

To give one example, in type~$D_4$ with the labeling
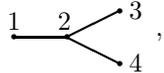
\[
\setlength{\unitlength}{1pt}
\begin{picture}(40,26)(0,0)
\thicklines
\put(0,10){\circle*{2}}
\put(20,10){\circle*{2}}
\put(40,0){\circle*{2}}
\put(40,20){\circle*{2}}
\put(0,10){\line(1,0){20}}
\put(20,10){\line(2,1){20}}
\put(20,10){\line(2,-1){20}}
\put(0,16){\makebox(0,0){$1$}}
\put(19,16){\makebox(0,0){$2$}}
\put(46,20){\makebox(0,0){$3$}}
\put(46,0){\makebox(0,0){$4$}}
\put(55,10){\makebox(0,0){,}}
\end{picture}
\]
we have
\[
F(\alpha_1+2\alpha_2+\alpha_3+\alpha_4)= u_1 u_3 u_4 + u_2^2 +
\sum_{1\leq i<j\leq 4} u_i u_j + \sum_{i\neq 2} u_i + 2u_2 +1.
\]

\proof
All we need to do is to verify that the polynomials given by
(\ref{eq:F-2-restricted}) (together with $F[-\alpha_i] = 1$ for all $i \in I$)
satisfy the relation (\ref{eq:tau-F}) in Theorem~\ref{th:F-polynomials}.

Let us consider a more general situation.
Let $I$ be the vertex set of an arbitrary finite bipartite graph
(without loops and multiple edges);
we will write $i \leftrightarrow j$ to denote that two vertices
$i, j \in I$ are adjacent to each other.
Let $Q$ be a free $\ZZ$-module with a chosen basis $(\alpha_i)_{i\in I}$.
A vector $\alpha = \sum_{i \in I} a_i \alpha_i \in Q$
is called \emph{2-restricted} if $0 \leq a_i \leq 2$ for all~$i\in I$.

\begin{lemma}
\label{lem:F-2-restricted-reduced}
Let $\alpha$ be a 2-restricted vector,
and let $F[\alpha]$ denote the polynomial
in the variables $u_i \ (i \in I)$ defined by {\rm(\ref{eq:F-2-restricted})}.
Then
\begin{equation}
\label{eq:F-2-restricted-transformed}
F[\alpha] = \sum_{\gamma} 2^{e(\gamma;\alpha)} u^\gamma
\prod_{j \in I_-}(u_j + 1)^{\max (a_j - \sum_{i \leftrightarrow j} c_i, 0)}
\ ,
\end{equation}
where the sum is over all $\alpha$-acceptable integer vectors
$\gamma = \sum c_i \alpha_i$
such that
\begin{equation}
\label{eq:funny-condition}
\text{$c_j = 0$ whenever $j\in I_-$ and $a_j > \sum_{i \leftrightarrow j}
c_i\,$.}
\end{equation}
\end{lemma}

\proof
The proof is based on regrouping the summands in
(\ref{eq:F-2-restricted}) according to the projection that
is defined on the set of $\alpha$-acceptable vectors $\gamma$ as
follows: it replaces each coordinate $c_j$ that violates the condition
(\ref{eq:funny-condition}) by~0.

The equivalence of (\ref{eq:F-2-restricted}) and
(\ref{eq:F-2-restricted-transformed}) is then verified as follows.
Suppose that $\gamma = \sum c_i \alpha_i$ is an
$\alpha$-acceptable integer vector.
Suppose furthermore that $j\in I_-$ is such that
$a_j>\sum_{i \leftrightarrow j} c_i\,$.
It is easy to check that, once the values of $a_j$ and $\sum_{i
  \leftrightarrow j} c_i$ have been fixed, the possible choices of
$c_j$ are determined as shown in the first three columns of
Table~\ref{table:case-by-case}.
Comparing the last two columns completes the verification.
\endproof

\begin{table}[hb]
\begin{center}
\begin{tabular}{ccccc}
$a_j$ & $\sum_{i \leftrightarrow j} c_i$ & $c_j$ &
\begin{tabular}{c}
replacing $c_j$ by 0 results in\\
dividing $2^{e(\gamma;\alpha)} u^\gamma$ by:
\end{tabular}
& $(u_j+1)^{\max(a_j - \sum_{i \leftrightarrow j} c_i, 0)}$
\\
\hline
1 & 0 & 0 & 1  & $1+u_j$\\
1 & 0 & 1 & $u_j$ \\
\hline
2 & 0 & 0 & 1 \\
2 & 0 & 1 & $2u_j$ & $(1+u_j)^2$\\
2 & 0 & 2 & $u_j^2$ \\
\hline
2 & 1 & 0 & 1 & $1+u_j$\\
2 & 1 & 1 & $u_j$ \\[.1in]
\end{tabular}
\end{center}

\smallskip

\caption{Proof of Lemma~\ref{lem:F-2-restricted-reduced}}
\label{table:case-by-case}
\end{table}

It will be convenient to restate
Lemma~\ref{lem:F-2-restricted-reduced} as follows.
For an integer vector $\gamma_+ = \sum_{i \in I_+} c_i \alpha_i$
satisfying the condition
\begin{equation}
\label{eq:pos-part-acceptable}
\text{$0 \leq c_i \leq a_i$ for all $i \in I_+\,$},
\end{equation}
we define  the polynomial
$H[\alpha:\gamma_+]$ in the variables $u_j \ (j \in I_-)$~by
\begin{equation}
\label{eq:H-2-restricted}
H[\alpha:\gamma_+] =  \sum_{\gamma_-} 2^{e(\gamma_+ + \gamma_-;\alpha)}
u^{\gamma_-} \ ,
\end{equation}
where the sum is over all vectors
$\gamma_- = \sum_{j \in I_-} c_j \alpha_j$ such that
$(\gamma_+ + \gamma_-)$ is $\alpha$-acceptable, and
$c_j = 0$ whenever $a_j > \sum_{i \leftrightarrow j} c_i$.
Then
\begin{equation}
\label{eq:F-2-restricted-broken-up}
F[\alpha] = \sum_{\gamma_+} u^{\gamma_+}
H[\alpha:\gamma_+]
\prod_{j \in I_-}(u_j + 1)^{\max (a_j - \sum_{i \leftrightarrow j} c_i, 0)}\ ,
\end{equation}
where the sum is over all integer vectors $\gamma_+ = \sum_{i
  \in I_+} c_i \alpha_i$ satisfying (\ref{eq:pos-part-acceptable}).

\begin{lemma}
\label{lem:H-2-restricted-tau}
Suppose that both $\alpha = \sum_{i \in I} a_i \alpha_i$ and $\tau_- \alpha$ are 2-restricted.
Denote $\alpha_+ = \sum_{i \in I_+} a_i \alpha_i$.
Then $H[\alpha:\gamma_+] = H[\tau_- \alpha:\alpha_+ - \gamma_+]$
for any integer vector $\gamma_+$ satisfying {\rm (\ref{eq:pos-part-acceptable})}.
\end{lemma}

\proof
We will first prove that the sets of monomials $u^{\gamma_-}$ that
contribute to $H[\alpha:\gamma_+]$ and $H[\tau_- \alpha:\alpha_+ -
\gamma_+]$ with positive coefficients are the same,
and then check the equality of their coefficients.
For the first task, we need to show that if integer vectors
$\alpha = \sum_{i \in I} a_i \alpha_i$ and
$\gamma = \sum_{i \in I} c_i \alpha_i$ satisfy the conditions
\begin{itemize}
\item[(0)] $0\leq a_i\leq 2$ for $i \in I$, and $0\leq -a_j +\sum_{i \leftrightarrow
    j} a_i\leq 2$ for $j \in I_-$;
\item[(1)] $0 \leq c_i \leq a_i$ for $i \in I$;
\item[(2)] $c_i+c_j \leq 2$ for any adjacent $i$ and~$j$;
\item[(3)] there is no simple path $(i_0, \dots, i_m)$, $m\geq 1$, 
with $c_0=\cdots=c_m=1$ and $a_0=a_m=1$;
\item[(4)] if $c_j>0$ and $j\in I_-\,$, then $a_j \leq \sum_{i \leftrightarrow j}
c_i\,$,
\end{itemize}
then the same conditions are satisfied for the vectors
$\tilde\alpha=\tau_- \alpha= \sum_{i \in I} \tilde a_i \alpha_i$
and $\tilde\gamma 
= \sum_{i\in I} \tilde c_i \alpha_i\,$, where
\begin{equation}
\label{eq:tilde-variables}
\begin{array}{ll}
\text{$\tilde a_i = a_i$ \ \ for $i \in I_+$;}\\[.1in]
\text{$\tilde a_j=-a_j+\sum_{i \leftrightarrow j} a_i$\ \  for $j\in I_-$;}
\end{array}
\qquad
\begin{array}{ll}
\text{$\tilde c_i = a_i-c_i$\ \  for $i \in I_+$;}\\[.1in]
\text{$\tilde c_j= c_j$\ \  for $j\in I_-$.}
\end{array}
\end{equation}
We do not have to prove the converse implication
since the map $(\alpha,\gamma)\mapsto (\tilde\alpha,\tilde\gamma)$
is involutive.

Here is the crucial part of the required verification: we claim
that conditions (0)-(4) leave only the five possibilities shown in
Table~\ref{table:5-cases} for the vicinity of a point $j \in I_-$ such
that $c_j > 0$.
(We omit the pairs of the form $(a_i,c_i)=(0,0)$ since they will be of
no relevance in the arguments below.)
To prove this, 
we first note that by (4), $c_i \geq 1$ for some $i \leftrightarrow j$.
It then follows from (2) that $c_j =1$, and furthermore $c_i \leq 1$
for all $i \leftrightarrow j$.
By (3), if $a_j = 1$, then $(a_i, c_i) \neq (1,1)$ for $i \leftrightarrow j$;
and if $a_j = 2$, then there is at most one index $i$ such that $i \leftrightarrow j$ and
$(a_i, c_i) = (1,1)$.
Combining (0), (1) and (4), we obtain the chain of inequalities
\[
1 \leq c_j \leq a_j \leq \textstyle\sum_{i \leftrightarrow j} c_i
\leq \textstyle\sum_{i \leftrightarrow j} a_i \leq a_j + 2
\]
which in particular imply that there can be at most two indices
$i$ such that $i \leftrightarrow j$ and $a_i > c_i$.
An easy inspection now shows that all these restrictions combined do
indeed leave only the five possibilities in Table~\ref{table:5-cases}.

\begin{table}[ht]
\begin{center}
\begin{tabular}{ccccccc}
$(a_j,c_j)$ & \quad $\bigl((a_i,c_i) \bigr)_{i \leftrightarrow j}$
&\quad  $(\tilde a_j, \tilde c_j)$ & \quad $\bigl((\tilde a_i,\tilde c_i) \bigr)_{i \leftrightarrow j}$
\\
\hline
(1,1) & $(2,1)$ & (1,1) & $(2,1)$\\
\hline
(1,1) & $(2,1)$, $(1,0)$ & (2,1) & $(2,1)$, $(1,1)$\\
\hline
(2,1) & $(2,1)$, $(1,1)$ &(1,1) & $(2,1)$, $(1,0)$ \\
\hline
(2,1) & $(2,1)$, $(2,1)$&  (2,1) & $(2,1)$, $(2,1)$\\
\hline
(2,1) & $(2,1)$, $(1,1)$, $(1,0)$& (2,1) & $(2,1)$, $(1,0)$, $(1,1)$ \\[.1in]
\end{tabular}
\end{center}
\caption{Proof of Lemma~\ref{lem:H-2-restricted-tau}}
\label{table:5-cases}
\end{table}


Conversely, if we assume that $\alpha$ and $\gamma$ satisfy (0)
and $0 \leq c_i \leq a_i$ for $i \in I_+$, then the restrictions
in Table~\ref{table:5-cases} imply the rest of (1) as well as (2)
and~(4).
These restrictions are evidently preserved under the
transformation (\ref{eq:tilde-variables}): compare the left and the
right halves of Table~\ref{table:5-cases}.
It only remains to show the following:
if $(\alpha, \gamma)$ and $(\tilde \alpha, \tilde \gamma)$ satisfy
(0), (1), (2), (4) and the
restrictions in Table~\ref{table:5-cases} but $(\tilde \alpha, \tilde
\gamma)$ violates
the only non-local condition (3), then
$(\alpha, \gamma)$ must also violate~(3).

Note that in each of the five cases, condition (3)
is satisfied by $(\tilde \alpha, \tilde \gamma)$ in the immediate vicinity of the vertex~$j$.
It follows that (3) could only be violated by
a path that has at least one non-terminal vertex $i\in I_+$
(then necessarily $\tilde a_i=2$ and $\tilde c_i=1$).
It is immediate from Table~\ref{table:5-cases} that the segments of
this path that lie between such vertices remain intact under the
involution $(\alpha,\gamma)\leftrightarrow (\tilde\alpha,\tilde\gamma)$; i.e.,
$a_i=\tilde a_i=2$ and $c_i=\tilde c_i=1$ holds throughout these
segments. The possibilities at the ends of the path are then examined
one by one with the help of Table~\ref{table:5-cases}; in each case,
we verify that condition (3) is violated by $(\alpha, \gamma)$, and we
are done.

To complete the proof of Lemma~\ref{lem:H-2-restricted-tau},
we need to check the equality of the coefficients in the polynomials
$H[\alpha:\gamma_+]$ and $H[\tau_- \alpha:\alpha_+ -\gamma_+]$.
In other words, we need to show that (0)--(4) implies
$e(\gamma;\alpha)=e(\tilde\gamma;\tilde\alpha)$.
(Recall that $e(\gamma;\alpha)$ was defined in
Proposition~\ref{pr:F-multiplicity-1-2}.)
For this, we note
that $a_i=2, c_i=1$
is equivalent to $\tilde a_i=2, \tilde c_i=1$ for $i\in I_+$;
and if a vertex $j\in I_-$ belongs to a connected component in
question, then we must be in the situation described in row~4 of
Table~\ref{table:5-cases}, so the transformation
$(\alpha,\gamma)\mapsto (\tilde\alpha,\tilde\gamma)$
does not change the vicinity of~$j$.
\endproof

With Lemmas~\ref{lem:F-2-restricted-reduced} and
\ref{lem:H-2-restricted-tau} under our belt, the proof of
Proposition~\ref{pr:F-multiplicity-1-2} is now completed as follows.
Interchanging $I_+$ and $I_-$ if necessary, it suffices to check
(\ref{eq:tau-F}) with $\varepsilon = +$.
Since (\ref{eq:F-2-restricted}) gives in particular
$F[\alpha_i] = u_i + 1$, (\ref{eq:tau-F}) checks for
$\alpha = \pm \alpha_i$, i.e., when one of $\alpha$ and
$\tau_- \alpha$ may be negative.
It remains to verify (\ref{eq:tau-F}) when $\varepsilon = +$,
both roots $\alpha$ and $\tau_- \alpha$ are positive,
and the polynomials $F[\alpha]$ and $F[\tau_- \alpha]$ are given by
(\ref{eq:F-2-restricted}).
Note also that when $\Phi$ is simply-laced, the dual root system
is canonically identified with $\Phi$ by a linear isomorphism of
ambient spaces; so we can replace each coefficient
$[\alpha^\vee : \alpha^\vee_i]$ appearing in (\ref{eq:tau-F})
by $[\alpha : \alpha_i] = a_i$.
Using 
(\ref{eq:F-2-restricted-broken-up}) and Lemma~\ref{lem:H-2-restricted-tau},
we obtain:
\begin{eqnarray*}
&& \Bigl({\prod_{i \in I_+} u_i^{a_i}}\Bigr)
\Bigl({\prod_{j \in I_-} (u_j + 1)^{-a_j}} \Bigr)
\,\tau_+ (F[\alpha])\\
 &&
=\sum_{\gamma_+} H[\alpha:\gamma_+]
\Bigl(\prod_{i \in I_+} u_i^{a_i - c_i}\Bigr)
\prod_{j \in I_-} (u_j + 1)^{\max (0, - a_j + \sum_{i \leftrightarrow
    j} c_i)}\\
&&
=\sum_{\gamma_+} H[\tau_- \alpha: \alpha_+ - \gamma_+]
\, u^{\alpha_+ - \gamma_+}
\prod_{j \in I_-} (u_j + 1)^{\max (\tilde a_j -
\sum_{i \leftrightarrow j} \tilde c_i, 0)}\\
&&=F[\tau_- \alpha]
\end{eqnarray*}
(here we used notation (\ref{eq:tilde-variables})).
Proposition~\ref{pr:F-multiplicity-1-2} is proved.
\endproof

\begin{remark}{\rm
We note that formula~(\ref{eq:F-2-restricted}) also holds for the Fibonacci
polynomial $F[\alpha]$ of an arbitrary 2-restricted root $\alpha$ in an
exceptional root system of type $E_6$, $E_7$, or~$E_8$.
The proof remains unchanged; the only additional ingredient is the
statement, easily verifiable by a direct computation, that any such root
can be obtained from a root of the form $-\alpha_i$, $i\in I$,
by a sequence of transformations $\tau_\pm$, so that all intermediate
roots are also 2-restricted.
}
\end{remark}

Formula~(\ref{eq:F-2-restricted}) becomes much simpler in the
special case when a positive root $\alpha$ is
\emph{multiplicity-free}, i.e., $[\alpha: \alpha_i] = 1$ for all $i$.
Let us denote
\begin{equation}
\label{eq:support}
{\rm Supp}(\alpha) = \{i \in I: [\alpha: \alpha_i] \neq 0\} \ .
\end{equation}
We call a subset $\Omega \subset I$ \emph{totally disconnected}
if $\Omega$ contains no two indices that are adjacent in the Coxeter graph.
As a special case of (\ref{eq:F-2-restricted}), we obtain the following.

\begin{proposition}
\label{pr:F-multiplicity-free}
For a multiplicity-free positive root $\alpha$, we have
\begin{equation}
\label{eq:F-mult-free}
F[\alpha] = \sum_{\Omega} \prod_{i \in \Omega} u_i \ ,
\end{equation}
where the sum is over all totally disconnected subsets
$\Omega \subset {\rm Supp}(\alpha)$.
\end{proposition}

\begin{example}
\label{example:fib-A}
{\rm
In the type~$A_n$ case, we can take $I = [1,n] = \{1, \dots, n\}$, with
$i$ and $j$ adjacent whenever $|i-j| = 1$.
Every  positive root is multiplicity-free, and their supports are
all intervals $[a,b] \subset [1,n]$.
Thus, we have
\begin{equation}
\label{eq:F-intervals}
F[\sum_{i \in [a,b]} \alpha_i] = F[a,b] =
\sum_{\Omega \subset [a,b]} \prod_{i \in \Omega} u_i \ ,
\end{equation}
the sum over totally disconnected subsets $\Omega \subset [a,b]$.
For example, for $n = 3$, the Fibonacci polynomials are given by
\begin{equation*}
\label{eq:A3-F}
\begin{array}{l}
F[1,1]=u_1\!+\!1,\\
F[2,2]=u_2\!+\!1,\\
F[3,3]=u_3\!+\!1,
\end{array}
{\ \ }
\begin{array}{l}
F[1,2]=u_1\!+\!u_2\!+\!1,\\
F[2,3]=u_2\!+\!u_3\!+\!1,
\end{array}
{\ \ }
F[1,3]= u_1u_3\!+\!u_1\!+\!u_2\!+\!u_3\!+\!1.
\end{equation*}
When all the $u_i$ are set to $1$, the polynomials $F[a,b]$ specialize to
Fibonacci numbers, which explains our choice of the name.
}
\end{example}

\medskip

\noindent {\bf Step 3. Exceptional types.}
To complete the proof of Theorem~\ref{th:F-polynomials},
it remains to consider the types $E_6$, $E_7$ and~$E_8\,$.
In each of these cases we used (\ref{eq:tau-F}) to recursively compute
all Fibonacci polynomials $F[\alpha]$
with the help of a \texttt{Maple} program.
Since some of the expressions involved are very large
(for example, for the highest root $\alpha_{\max}$ in type~$E_8$,
the polynomial $F[\alpha_{\max}](u_1,\dots,u_8)$ has 26908 terms in its
monomial expansion, and its largest coefficient is~3396),
one needs an efficient way to organize these computations.

We introduce the variables $v_i=u_i+1$, for $i\in I$.
Suppose that  the polynomial $F[\alpha]$, for some root
$\alpha\in\Phi_{\geq-1}$, has already been computed.
(As initial values, we can take $\alpha=-\alpha_i$, $i\in I$,
with $F[-\alpha_i]=1$.)
For a sign~$\varepsilon$,
let us express $F[\alpha]$ as a polynomial in the variables
\begin{equation}
\label{eq:uv-variables}
(u_i : i\in I_\varepsilon) \cup (v_i : i\in
  I_{-\varepsilon}).
\end{equation}
In these variables, $\tau_\epsilon$ becomes a (Laurent) monomial transformation;
in particular, it does not change the number of terms in the monomial
expansion of~$F[\alpha]$.
Using the recursion (\ref{eq:tau-F}), rewritten in the form
\begin{equation}
\label{eq:tau-F-uv}
F[\tau_{- \varepsilon} (\alpha)]=
\frac{\prod_{\varepsilon (i) = \varepsilon}
u_i^{[\alpha^\vee : \alpha^\vee_i]}}
{\prod_{\varepsilon (i) = - \varepsilon}
v_i^{[\alpha^\vee : \alpha^\vee_i]}}
\cdot
\tau_\varepsilon (F[\alpha]),
\end{equation}
we compute $F[\tau_{- \varepsilon} (\alpha)]$
as a function, indeed a polynomial, in the variables~(\ref{eq:uv-variables}).
We then make the substitution
$v_i\leftarrow u_i+1$ for all $i\in I_{-\varepsilon}$ to express
 $F[\tau_{- \varepsilon} (\alpha)]$ in terms of the original variables
$(u_i)_{i\in I}$, and record the result in our files.
Next, we substitute
$u_i\leftarrow v_i-1$ for all $i\in I_{\varepsilon}$,
thus expressing $F[\tau_{- \varepsilon} (\alpha)]$ as a polynomial in
the variables
\begin{equation*}
\label{eq:other-uv-variables}
(u_i : i\in I_{-\varepsilon}) \cup (v_i : i\in
  I_{\varepsilon}).
\end{equation*}
We then reset $\varepsilon:=-\varepsilon$ and $\alpha:=\tau_{- \varepsilon}
(\alpha)$, completing the loop.
The steps described above in this paragraph are repeated until we
arrive at $\alpha=-\alpha_i$, for some $i\in I$.
Taking as initial values for~$\alpha$ all possible negative simple
roots,
we can compute all polynomials~$F[\alpha]$, $\alpha\in \Phi_{\geq-1}$.

To check the validity of Theorem~\ref{th:F-polynomials}
for a given root system~$\Phi$, we need to
verify that
\begin{itemize}
\item
every $F[\alpha]$ is a polynomial;
\item
when expressed in the variables $(u_i)_{i\in I}$, each $F[\alpha]$ has
nonnegative integer coefficients and constant term $1$;
\item
each time the process arrives at $\alpha=-\alpha_i$,
it returns $F[\alpha]=1$.
\end{itemize}
The algorithm described above does indeed verify these properties
for the types $E_6$, $E_7$ and $E_8$.
This completes the proof of Theorem~\ref{th:F-polynomials}.

\section{Generalized assohiahedra}
\label{sec:cluster-complexes}

Throughout this section, we retain the terminology and notation from the
previous sections, with one important exception:
we drop the assumption that the Cartan matrix $A$ is indecomposable.
Thus, the corresponding (reduced finite) root system $\Phi$ is no
longer assumed to be irreducible, and its Coxeter graph can be a
forest, rather than a tree.
We are in fact forced to pass to this more general setting
because most of the proofs in this section are based on passing from
$\Phi$ to a proper subsystem of $\Phi$ which may not be
irreducible even if $\Phi$~is.
For every subset $J \subset I$, let $\Phi(J)$ denote the root
subsystem in $\Phi$ spanned by the set of simple roots
$\{\alpha_i: i \in J\}$.
If $I_1, \dots, I_r$ are the connected components of~$I$,
then $\Phi$ is the disjoint union of irreducible root systems
$\Phi (I_1), \dots, \Phi (I_r)$, and all results of the
previous sections extend in an obvious way to this more general setting.
In particular, we can still subdivide $I$ into the disjoint union
of two totally disconnected subsets $I_+$ and $I_-$ (by doing this
independently for each connected component of $I$), and consider
the corresponding piecewise-linear involutions $\tau_+$ and
$\tau_-$ of the set $\Phi_{\geq -1}$.
Theorem~\ref{th:Y-Phi} holds verbatim.
We note that if
$\alpha \in \Phi_{\geq -1}$ belongs to an irreducible subsystem
$\Phi (I_k)$, then the corresponding Laurent polynomial
$Y[\alpha]$ involves only variables $u_i$ for $i \in I_k$.

\subsection{The compatibility degree}
\label{sec:compatibility}
We define the function
\[
\begin{array}{ccc}
\Phi_{\geq -1} \times \Phi_{\geq -1} &\to& \ZZ_{\geq 0}\\[.1in]
(\alpha,\beta) &\mapsto & (\alpha \| \beta)\,,
\end{array}
\]
called the \emph{compatibility degree} (of $\alpha$ and~$\beta$), by
\[
(\alpha \| \beta) = [Y[\alpha] + 1]_{\rm trop} (\beta)
\]
(cf.\ (\ref{eq:compatibility-degree})).
As before,
this notation means evaluating the ``tropical specialization" of the Laurent
polynomial $Y[\alpha] + 1$ (obtained by replacing ordinary addition and
multiplication by the operations (\ref{eq:max-+})) at the tuple $(u_i
= [\beta:\alpha_i])_{i \in I}$.
By virtue of its definition, this function is uniquely
characterized by the following two properties:
\begin{eqnarray}
\label{eq:compatibility-1}
&(- \alpha_i \| \beta) = \max \ ([\beta: \alpha_i], 0) , &
\\
\label{eq:compatibility-2}
&(\tau_\varepsilon \alpha \| \tau_\varepsilon \beta) = (\alpha \|
\beta) , &
\end{eqnarray}
for any $\alpha, \beta \in \Phi_{\geq -1}$, any $i\in I$,
and any sign~$\varepsilon$.

The next proposition gives an unexpectedly simple explicit
formula for the compatibility degree.
This formula involves the perfect bilinear
pairing
\[
\begin{array}{ccc}
Q^\vee \times Q &\to& \ZZ \\
(\xi,\gamma) &\mapsto & \{\xi, \gamma\}
\end{array}
\]
(recall that $Q$ is the root lattice, and $Q^\vee$ is the root lattice for the dual root system)
defined by
\begin{equation}
\label{eq:eps-pairing}
\{\xi, \gamma\} = \sum_{i \in I} \varepsilon (i) [\xi : \alpha_i^\vee]
[\gamma: \alpha_i]\ .
\end{equation}

\begin{proposition}
\label{pr:compatibility-explicit}
The compatibility degree $(\alpha \| \beta)$ is given by
\begin{equation}
\label{eq:compatibility-explicit}
(\alpha \| \beta) = \max(\{\alpha^\vee, \tau_+ \beta\},
\{\tau_+ \alpha^\vee, \beta\}, 0)\,.
 \end{equation}
Alternatively,
\begin{equation}
\label{eq:compatibility-change-of-sign}
(\alpha \| \beta)
=\max(- \{\tau_- \alpha^\vee, \beta\}, - \{\alpha^\vee, \tau_- \beta\}, 0)
 \ .
 \end{equation}
\end{proposition}

\proof
First let us show that (\ref{eq:compatibility-explicit})
and (\ref{eq:compatibility-change-of-sign}) agree with each other,
i.e., define the same function
$\Phi_{\geq -1} \times \Phi_{\geq -1} \to \ZZ_{\geq 0}$.
To do this, we note that the pairing $\{\cdot,\cdot\}$ 
satisfies the identity
\begin{equation}
\label{eq:t-adjoint}
\{\xi, t_\varepsilon \gamma\} =
- \varepsilon \Biggl(\sum_{i \in I} \xi_i \gamma_i
+\sum_{\varepsilon (i)
= \varepsilon = - \varepsilon (j)} a_{ij}
\xi_i \gamma_j\Biggr)
= - \{t_{- \varepsilon} \xi, \gamma\}
\end{equation}
for any sign $\varepsilon$, any $\xi\in Q^\vee$ and any $\gamma\in Q$,
where we abbreviate
$\xi_i = [\xi : \alpha_i^\vee]$ and $\gamma_i = [\gamma : \alpha_i]$
(this follows from (\ref{eq:eps-pairing}) and
(\ref{eq:tau-pm-tropical})).
Since $t_\pm$ agrees with $\tau_\pm$ on positive roots
and coroots (see Proposition~\ref{pr:tau-t}.2), we conclude that
(\ref{eq:compatibility-explicit})
and (\ref{eq:compatibility-change-of-sign}) agree with each other
on $\Phi_{> 0} \times \Phi_{> 0}$.
It remains to treat the case when at least one of $\alpha$ and
$\beta$ belongs to $- \Pi$.
If say $\alpha = - \alpha_i$ with $i \in I_+$ then we have
$$\{\alpha^\vee, \tau_+ \beta\} = [\beta: \alpha_i] +
\sum_{j \neq i} a_{ij} \max ([\beta: \alpha_j],0) \leq
[\beta:\alpha_i] = \{\tau_+ \alpha^\vee, \beta\}
$$
and
$$- \{\alpha^\vee, \tau_- \beta\} =
[\beta:\alpha_i] = - \{\tau_- \alpha^\vee, \beta\};$$
thus, (\ref{eq:compatibility-explicit})
and (\ref{eq:compatibility-change-of-sign}) agree in this case as well.
The cases when $\alpha = - \alpha_i$ with $i \in I_-$, or $\beta \in - \Pi$
are handled in the same way.

To complete the proof of Proposition~\ref{pr:compatibility-explicit},
note first that both (\ref{eq:compatibility-explicit})
and (\ref{eq:compatibility-change-of-sign}) agree with
(\ref{eq:compatibility-1}) (we just demonstrated this for
$\alpha = - \alpha_i$ with $i \in I_+$; the case $i \in I_-$ is
totally similar).
On the other hand, (\ref{eq:compatibility-explicit})
(resp., (\ref{eq:compatibility-change-of-sign})) makes it obvious
that $(\alpha \| \beta)$ is $\tau_+$-invariant
(resp., $\tau_-$-invariant), and we are done.
\endproof

\begin{remark}
{\rm Comparing (\ref{eq:compatibility-explicit})
and (\ref{eq:compatibility-change-of-sign}), we see that the
compatibility degree
$(\alpha \| \beta)$ does not depend on the choice of the sign
function $\varepsilon: I \to \{\pm 1\}$.
}
\end{remark}

The following proposition summarizes some properties of
$(\alpha \| \beta)$.

\pagebreak[2]

\begin{proposition}
\label{pr:compatibility-symmetry}
{\ }
\begin{enumerate}
\item
We have $(\alpha \| \beta) = (\beta^\vee \| \alpha^\vee)$
for every $\alpha, \beta \in \Phi_{\geq -1}$.\\
In particular, if $\Phi$ is simply-laced, then
$(\alpha \| \beta) = (\beta \| \alpha)$.

\item
If $(\alpha \| \beta) = 0$, then
$(\beta \| \alpha) = 0$.

\item
If $\alpha$ and $\beta$ belong
to $\Phi(J)_{\geq -1}$ for some proper subset $J \subset I$
then their compatibility degree with
respect to the root subsystem $\Phi(J)$ is equal to $(\alpha \| \beta)$.
\end{enumerate}
\end{proposition}

\proof
Parts (1) and (3) are immediate from
(\ref{eq:compatibility-explicit});
to verify (3) in the only nontrivial case where both $\alpha$ and
$\beta$ are positive roots, expand the terms in
(\ref{eq:compatibility-explicit}) using (\ref{eq:t-adjoint}).
To show (2), recall the following well known property of root
systems: there is a linear isomorphism between the dual root
lattices $Q$ and $Q^\vee$ under which every coroot $\alpha^\vee$
becomes a positive rational multiple of the corresponding root $\alpha$.
The definition (\ref{eq:eps-pairing}) implies that under this
identification, $\{\cdot,\cdot\}$ becomes a \emph{symmetric}
bilinear form on $Q$.
It follows that $\{\alpha^\vee, \tau_+ \beta\}$ and
$\{\tau_+ \beta^\vee, \alpha\}$ (resp., $\{\tau_+  \alpha^\vee, \beta\}$
and $\{\beta^\vee, \tau_+ \alpha\}$) are of the same sign.
In view of (\ref{eq:compatibility-explicit}), we conclude that
$$(\alpha \| \beta) = 0 \Leftrightarrow
\max(\{\alpha^\vee, \tau_+ \beta\}, \{\tau_+ \alpha^\vee, \beta\})
\leq 0$$
$$\Leftrightarrow
\max(\{\tau_+ \beta^\vee, \alpha\}, \{\beta^\vee, \tau_+ \alpha\})
\leq 0 \Leftrightarrow (\beta \| \alpha) = 0,$$
as claimed.
\endproof

\subsection{Compatible subsets and clusters}
\label{sec:compatible-subsets}

We say that two roots $\alpha,\beta\in\Phi_{\geq -1}$
are \emph{compatible} if $(\alpha \| \beta) = 0$.
In view of Proposition~\ref{pr:compatibility-symmetry}.2, the
compatibility relation is symmetric.
By (\ref{eq:compatibility-2}), both $\tau_+$ and $\tau_-$
preserve compatibility.

\begin{definition}
\label{def:clusters}
{\rm A subset of $\Phi_{\geq -1}$ is called \emph{compatible}
if it consists of mutually compatible elements.
The (root) \emph{clusters} associated to a root system $\Phi$ are the
maximal (by inclusion) compatible subsets of $\Phi_{\geq -1}$.}
\end{definition}

The following proposition will allow us to establish properties of
compatible subsets and clusters using induction on the rank $n = |I|$ of the root system.

\begin{proposition}
\label{pr:cluster-reduction}
{\ }
\begin{enumerate}
\item
Both $\tau_+$ and $\tau_-$ take compatible subsets to compatible subsets
and clusters to clusters.

\item
If $I_1, \dots, I_r\subset I$ are the connected
components of the Coxeter graph, then the compatible subsets (resp., clusters)
for $\Phi$ are the disjoint unions
$C_1 \bigsqcup \cdots \bigsqcup C_r$, where each $C_k$ is a
compatible subset (resp., cluster) for $\Phi (I_k)$.

\item
For every $i \in I$, the correspondence $C \mapsto C - \{-\alpha_i\}$ is a bijection
between the set of all compatible subsets (resp., clusters) for $\Phi$
that contain $- \alpha_i$ and the set of all compatible subsets
(resp., clusters)
for $\Phi (I - \{i\})$.
\end{enumerate}
\end{proposition}

\proof
Part~1 follows from the fact that $\tau_+$ and $\tau_-$
preserve compatibility.
Part~2 follows from the fact that $\tau_+$ and $\tau_-$ preserve
each set $\Phi (I_k)_{\geq -1}\,$.
Part~3 follows from~(\ref{eq:compatibility-1}).
\endproof

For a compatible subset $C$, we set
$$S_- (C) = \{i \in I: - \alpha_i \in C\} \ ,$$
and call $S_- (C)$ the \emph{negative support} of $C$.
We say that $C$ is \emph{positive} if $S_- (C) = \emptyset$, i.e.,
if $C$ consists of positive roots only.
The following proposition is obtained by iterating
Proposition~\ref{pr:cluster-reduction}.3.

\begin{proposition}
\label{pr:reduction-to-positivity}
For every subset $J \!\subset\! I$,
the correspondence $C\! \mapsto\! C\!-\!\{-\alpha_i\!:\!i\!\in\!J\}$ is a bijection
between the set of all compatible subsets (resp., clusters) for $\Phi$
with negative support $J$ and the set of all positive compatible
subsets (resp., clusters)
for $\Phi (I - J)$.
\end{proposition}

We are now ready to prove the purity property for clusters.

\medskip

\noindent {\bf Proof of Theorem~\ref{th:cluster-purity}.}
We need to show that every cluster $C$ for $\Phi$ is a $\ZZ$-basis of the
root lattice~$Q$.
In view of Proposition~\ref{pr:reduction-to-positivity}, it suffices to
prove this in the case when $C$ is positive.
Combining Propositions~\ref{pr:cluster-reduction}.1 and
\ref{pr:tau-t}.2, we see that both collections $t_+ (C)$
and $t_- (C)$ are also clusters.
Obviously, each of them is a $\ZZ$-basis of $Q$ if and only if
this is true for $C$.
Iterating this construction if necessary, we will arrive at a
cluster $C'$ which is no longer positive; this follows from
Theorem~\ref{th:dihedral}.1.
Now it suffices to prove that $C'$ is a $\ZZ$-basis of $Q$.
Again using Proposition~\ref{pr:reduction-to-positivity},
it is enough to show this for the positive part of $C'$, which is
a cluster for a root subsystem of smaller rank.
Induction on the rank completes the proof.
\endproof

\subsection{Counting compatible subsets and clusters}

Let $\Phi$ be a root system of rank~$n$.
For $k = 0, \dots, n$, let $f_k (\Phi)$ denote the number of
compatible $k$-subsets of $\Phi_{\geq -1}$;
in particular, $f_n (\Phi)$ is the number of clusters associated to $\Phi$.
We have $f_0 (\Phi) = 1$, and
$f_1 (\Phi) = |\Phi_{\geq -1}|$; if $\Phi$ is irreducible, then the
latter number is equal to $|\Phi_{\geq -1}| = n(h+2)/2$, where
$h$ is the Coxeter number of $\Phi$.
Let
$$f(\Phi) = \sum_{0 \leq k \leq n} f_k (\Phi) x^k$$
be the corresponding generating function.

\begin{proposition}
\label{pr:f-vector}
{\ }
\begin{enumerate}
\item
We have
$f (\Phi_1 \times \Phi_2) = f(\Phi_1) f(\Phi_2)$.

\item
If $\Phi$ is irreducible, then, for every
$k \geq 1$, we have
$$f_k (\Phi) =
\frac{h+2}{2k} \sum_{i \in I} f_{k-1}(\Phi (I - \{i\})) \,.$$
Equivalently,
$$\frac{d f(\Phi)}{d x} =
\frac{h+2}{2} \sum_{i \in I} f(\Phi (I - \{i\})) \ .$$
\end{enumerate}
\end{proposition}

\proof
Part~1 follows at once from Proposition~\ref{pr:cluster-reduction}.2.
To prove Part~2, we count in two different ways
the number of pairs $(\alpha,S)$, where $S \subset \Phi_{\geq -1}$
is a compatible $k$-subset, and $\alpha \in S$.
On one hand, the number of pairs in question is $k f_k(\Phi)$.
On the other hand, combining Proposition~\ref{pr:tau
  orbits}/Theorem~\ref{th:dihedral},
formula (\ref{eq:compatibility-1}), and
Proposition~\ref{pr:cluster-reduction} (Parts~1 and~3),
we conclude that the roots $\alpha$ belonging to each $D$-orbit
$\Omega$ in $\Phi_{\geq -1}$ contribute
$$\frac{h+2}{2} \sum_{i \in I: - \alpha_i \in \Omega} f_{k-1}(\Phi (I
- \{i\}))$$
to the count, implying the claim.
\endproof

Proposition~\ref{pr:f-vector} provides a recursive way to compute the
numbers~$f_k(\Phi)$.
It can also be used to obtain explicit formulas for $f_n (\Phi)$,
the total number of clusters.
When $\Phi$ is of some Cartan-Killing type $X_n$, we shall write $N(X_n)$
instead of $f_n(\Phi)$.

\begin{proposition}
\label{pr:number-of-clusters}
For every irreducible root system $\Phi$, say of type~$X_n\,$,
the corresponding number of clusters $N(X_n)$ is given in
Table~\ref{tab:cluster-numbers}.
\end{proposition}

\begin{table}[ht]
\begin{center}
\begin{tabular}{|c|c|c|c|c|c|c|c|c|c|}
\hline
$X_n$ & $A_n$ & $B_n,C_n$ & $D_n$ & $E_6$ & $E_7$ & $E_8$ & $F_4$  & $G_2$ \\
\hline
&&&&&&&&\\[-.1in]
$N (X_n)$ & 
$\frac{1}{n+2} \binom{2n+2}{n+1}$
& $\binom{2n}{n}$ 
&   $\frac{3n-2}{n}\binom{2n-2}{n-1}$ 
&
833
&
4160
&
25080
&
105
& $8$ \\[.05in]
\hline
\end{tabular}
\end{center}
\medskip
\caption{Counting the clusters}
\label{tab:cluster-numbers}
\end{table}

\vspace{-.1in}

Since the number of clusters in type $A_n$ is a Catalan number,
the numbers $N(X_n)$ can be regarded as generalizations of the
Catalan numbers to arbitrary Dynkin diagrams.
Another generalization appears in
Proposition~\ref{pr:N-positive}/Table~\ref{tab:cluster-numbers-positive}
below.

\proof
We will present the proof for the type~$A_n$; other types are treated
in a totally similar way.
Let us abbreviate $a_n = N(A_n)$.
We need to show that $a_n = c_{n+1}$, where
 $c_k = \frac{1}{k+1} \binom{2k}{k}$ denotes the $k$th Catalan number.

Proposition~\ref{pr:f-vector} produces the recursion
\begin{equation}
\label{eq:A-recursion}
a_n = \frac{n+3}{2n} \sum_{i=1}^{n} a_{i-1} a_{n-i} \,,
\end{equation}
for $n \geq 1$.
All we need to do is to check that the sequence $a_n=c_{n+1}$ satisfies
(\ref{eq:A-recursion}), together with the
initial condition $a_0 = 1$. In other words, we need to show that the
Catalan numbers satisfy
\[
c_{n+1} =
\frac{n+3}{2n} \sum_{i = 1}^{n} c_i c_{n+1-i}
\,.
\]
With the help of the well known identity $\sum_{i = 0}^{n+1} c_i c_{n+1-i}
=c_{n+2}$,
this is  transformed into
$c_{n+1}=\frac{n+3}{2n}(c_{n+2}-2c_{n+1})$, which is easily checked
using the formula for~$c_n$.
\endproof

\noindent {\bf Proof of Theorem~\ref{th:N-thru-exponents}.}
Formula (\ref{eq:N-thru-exponents}) follows from
Proposition~\ref{pr:number-of-clusters} by a case by case
inspection (the definition of the exponents $e_i$ and their
values for all irreducible root systems can be found in
\cite{bourbaki}).
\endproof

The appearance of exponents in the formula (\ref{eq:N-thru-exponents}) for the number of
clusters is a mystery to us at the moment.
To add to this mystery, a similar expression can
be given for the number $N^+(X_n)$ of \emph{positive} clusters.

\begin{proposition}
\label{pr:N-positive}
For every Cartan-Killing type $X_n$, the number of positive clusters is
given by
\begin{equation}
\label{eq:N-positive}
N^+(X_n) = \prod_{i=1}^n \frac{e_i + h - 1}{e_i + 1} \ ,
\end{equation}
where $e_1, \dots, e_n$ are the exponents of the root system of
type $X_n$, and $h$ is the Coxeter number.
Explicit formulas are given in
Table~\ref{tab:cluster-numbers-positive}.
\end{proposition}

\begin{table}[ht]
\begin{center}
\begin{tabular}{|c|c|c|c|c|c|c|c|c|c|}
\hline
$X_n$ & $A_n$ & $B_n,C_n$ & $D_n$ & $E_6$ & $E_7$ & $E_8$ & $F_4$  & $G_2$ \\
\hline
&&&&&&&&\\[-.1in]
$N (X_n)$ &
$\frac{1}{n+1} \binom{2n}{n}$
& $\binom{2n-1}{n}$
&   $\frac{3n-4}{n}\binom{2n-3}{n-1}$
&
418
&
2431
&
17342
&
66
&
5
\\[.05in]
\hline
\end{tabular}
\end{center}
\medskip
\caption{Counting the positive clusters}
\label{tab:cluster-numbers-positive}
\end{table}

\vspace{-.1in}

\proof
For each subset $J\! \subset\! I$, let $N(J)$
(resp., $N^+(J)$) denote the number of clusters (resp., positive clusters)
for the root system $\Phi (J)$.
By Proposition~\ref{pr:reduction-to-positivity}, we have
\begin{equation}
\label{eq:N(I)}
N(I) = \sum_{J \subset I} N^+(J) \,.
\end{equation}
By the inclusion-exclusion principle, this implies
\begin{equation}
\label{eq:N^+(I)}
N^+(I) = \sum_{J \subset I}(-1)^{|I-J|} N(J) \, .
\end{equation}
Substituting into the right-hand side the data from
Table~\ref{tab:cluster-numbers}, we can calculate $N^+ (X_n)$
for all types, and verify the
formula (\ref{eq:N-positive}) by a case by case inspection.
To illustrate, consider the case of type~$A_n$ (other cases are treated
in a similar way).
In that case, we can take $I=[1,n]$, and (\ref{eq:N^+(I)}) becomes
\begin{equation}
\label{eq:N^+(I)-type-A}
N^+(I)
= \sum_{J \subset [1,n]}(-1)^{|I-J|} c_J \, ,
\end{equation}
where $c_J$ denotes the product of the Catalan numbers $c_{|J_i|+1}$
over all connected components $J_i$ of~$J$.
Let us encode each $J\subset [1,n]$, say of cardinality
$n-k$, by a sequence $(j_0,\dots,j_k)$ of positive integers adding up
to~$n+1$, as follows:
\[
[1,n]-J = \{\,j_0\,,\,j_0+j_1\,,\,\dots\,,\,j_0+\cdots +j_{k-1}\,\}.
\]
Then $c_J=c_{j_0}\cdots c_{j_k}\,$, and therefore
(\ref{eq:N^+(I)-type-A}) can be rewritten as
\[
N^+ ([1,n]) = \sum_{k=0}^n (-1)^k
\sum_{\begin{array}{c}\\[-.2in]
\scriptstyle j_0+\dots+j_k=n+1\\[-.03in]
\scriptstyle j_i>0\end{array}}
c_{j_0} \cdots c_{j_k} \,.
\]
Let $C = C(t) = \sum_{j \geq 0} c_j t^j=1+t+2t^2+\cdots$ be the generating function for
the Catalan numbers; it is uniquely determined by the equation
$C = 1 + tC^2$.
We have
\[
\sum_{n \geq 0} N^+ ([1,n]) t^n = - t^{-1} \sum_{k \geq 0} (1 - C)^{k+1}
= \frac{C-1}{tC} = C\,;
\]
so $N^+ ([1,n])=c_n\,$, as needed.
\endproof



Needless to say, it would be nice to find a unified explanation of
(\ref{eq:N-thru-exponents}) and (\ref{eq:N-positive}).

\subsection{Cluster expansions}

\begin{definition}
{\rm A \emph{cluster expansion} of a vector $\gamma$ in the root
  lattice~$Q$
is a way to express $\gamma$ as
\begin{equation}
\label{eq:cluster-expansion}
\gamma = \sum_{\alpha \in \Phi_{\geq -1}} m_\alpha \alpha \, ,
\end{equation}
where all $m_\alpha$ are nonnegative integers, and
$m_\alpha m_\beta = 0$ whenever $\alpha$ and $\beta$ are not
compatible.
In other words, a cluster expansion is an expansion into a sum
of pairwise compatible roots in~$\Phi_{\geq -1}\,$.
}
\end{definition}


\begin{theorem}
\label{th:cluster-lattice}
Every element of the root lattice has a unique cluster expansion.
\end{theorem}

\proof
Our proof follows the same strategy as the proof of
Theorem~\ref{th:cluster-purity} given in
Section~\ref{sec:compatible-subsets}, although this time,
we need a little bit more preparation.
For $\gamma \in Q$, set
\[
\begin{array}{l}
S_+ (\gamma) = \{i \in I: [\gamma:\alpha_i] > 0\},\\[.1in]
S_- (\gamma) = \{i \in I: [\gamma:\alpha_i] < 0\} \,.
\end{array}
\]
In particular, for a positive root~$\alpha$, we have $S_+(\alpha)={\rm
  Supp}(\alpha)$ (cf.\ (\ref{eq:support})).

The following lemma is an easy consequence of
(\ref{eq:compatibility-1}).

\begin{lemma}
\label{lem:S-1}
Suppose that $\alpha\! \in\! \Phi_{\geq -1}$ occurs in a cluster expansion
of $\gamma$, that~is, $m_\alpha > 0$ in {\rm (\ref{eq:cluster-expansion})}.
Then either $\alpha$ is a positive root with
${\rm Supp} (\alpha) \subset S_+ (\gamma)$, or else
$\alpha = - \alpha_i$ for some $i \in S_- (\gamma)$.
In particular, if $\gamma \in Q_+\,$, then a cluster expansion of
$\gamma$ may only involve positive roots.
\end{lemma}

Let us denote
$$\gamma^{(+)} = \sum_{i \in I} \max([\gamma:\alpha_i],0) \alpha_i
= \sum_{i \in S_+(\gamma)} [\gamma:\alpha_i] \alpha_i \ .$$
The next lemma follows at once from Lemma~\ref{lem:S-1}.

\begin{lemma}
\label{lem:S-2}
A vector $\gamma \in Q$ has a unique cluster expansion if and only if
$\gamma^{(+)}$ has a unique cluster expansion with respect to the root
system $\Phi (S_+(\gamma))$.
\end{lemma}

In view of Lemma~\ref{lem:S-2}, it suffices to prove
Theorem~\ref{th:cluster-lattice} for $\gamma \in Q_+$.
Lemma~\ref{lem:S-1} implies in particular that the statement
holds for $\gamma = 0$, so we can assume that $\gamma \neq 0$.
We can also assume without loss of generality that
$\Phi$ is irreducible
(cf.\ Proposition~\ref{pr:cluster-reduction}.2).
Let $\varepsilon \in \{+,-\}$.
Combining  Propositions~\ref{pr:tau-t}.2 and
\ref{pr:cluster-reduction}.1, we conclude that
$$\gamma = \sum_{\alpha \in \Phi_{\geq -1}} m_\alpha \alpha$$
is a cluster expansion of $\gamma$ if and only if
$$t_\varepsilon \gamma = \sum_{\alpha \in \Phi_{\geq -1}} m_\alpha
\tau_\varepsilon \alpha$$
is a cluster expansion of $t_\varepsilon \gamma$.
Thus, $\gamma$ has a unique cluster expansion if and only if
$t_\varepsilon \gamma$ has.

To complete the proof, note that some product of the
transformations $t_+$ and $t_-$ must move $\gamma$ outside~$Q_+\,$.
Indeed, $w_\circ \gamma \in - Q_+$, and $w_\circ$ can
be written as such a product by Lemma~\ref{lem:lusztig-rw}.
Using this fact, we can assume without loss of generality that
already say $t_+ \gamma \notin Q_+$ (while $\gamma\in Q_+$).
By Lemma~\ref{lem:S-2}, the statement that $\gamma$ has a unique
cluster expansion follows from the same property for the vector $(t_+
\gamma)^{(+)}$, which lies in a root lattice of smaller rank.
The proof of Theorem~\ref{th:cluster-lattice} is now completed by
induction on the rank.
\endproof

\noindent {\bf Proof of Theorem~\ref{th:cluster-fan}.}
Theorem~\ref{th:cluster-fan} is essentially a direct corollary of
Theorem~\ref{th:cluster-lattice}.
We need to show two things:
\pagebreak[2]
\begin{itemize}
\item[(i)] no two of the cones $\RR_{\geq 0} C$ generated by clusters
have a common interior point;
\item[(ii)]
the union of these cones is the entire space $Q_\RR$.
\end{itemize}
To show (i), assume on the contrary that two of the cones
$\RR_{\geq 0} C$ have a common interior point.
Since the rational vector space $Q_\QQ$ is dense in $Q_\RR$, it
follows that there is a common interior point in $Q_\QQ$.
Multiplying if necessary by a suitable positive integer, we
conclude that there is also a common interior point in~$Q$.
Since we have already proved that every cluster is a $\ZZ$-basis
of $Q$, the latter statement contradicts the uniqueness of a
cluster expansion in Theorem~\ref{th:cluster-lattice}.

To show (ii), note that the existence of a
cluster expansion in Theorem~\ref{th:cluster-lattice}
implies that the union of the cones $\RR_{\geq 0} C$ contains~$Q$.
Since this union is closed in $Q_\RR$ and is stable under
multiplication by positive real numbers, it is the entire space
$Q_\RR$, and we are done.
\endproof

\subsection{Compatible subsets and clusters for the classical types}
\label{sec:clusters-classical}
{\ }

\noindent {\bf Type $A_n\,$.}
We use the standard labeling of the simple
roots by the set $I = [1,n] = \{1, \dots, n\}$.
Thus, the Coxeter graph is the
chain with the vertices $1,\dots,n$,
and the positive roots are given by
$\alpha_{ij} = \alpha_i + \alpha_{i+1} + \cdots + \alpha_j\,$,
for $1 \leq i \leq j \leq n$.
For $i \in [1,n]$, we set $\varepsilon (i) = (-1)^{i-1}$.

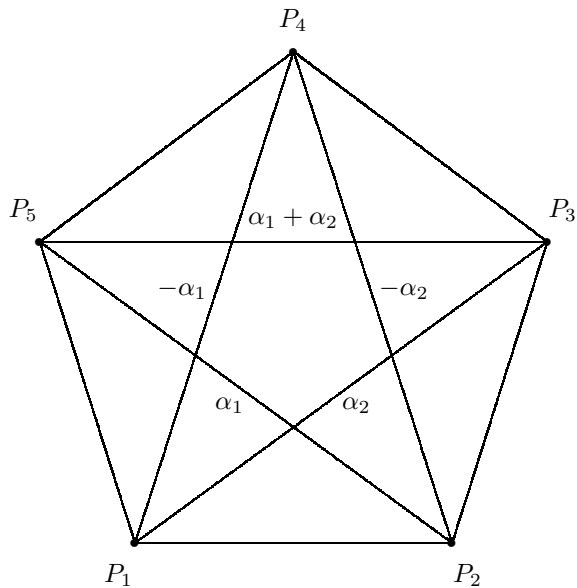
\begin{figure}[ht]
\begin{center}
\setlength{\unitlength}{6pt}
\begin{picture}(40,37)(-4,-2)
  \put(6,0){\line(1,0){20}}
  \put(0,19){\line(1,0){32}}
  \qbezier(6,0)(11,15.5)(16,31)
  \qbezier(26,0)(21,15.5)(16,31)
  \qbezier(6,0)(3,9.5)(0,19)
  \qbezier(26,0)(29,9.5)(32,19)
  \qbezier(0,19)(13,9.5)(26,0)
  \qbezier(32,19)(19,9.5)(6,0)
  \qbezier(0,19)(8,25)(16,31)
  \qbezier(32,19)(24,25)(16,31)

  \put(6,0){\circle*{.5}}
  \put(26,0){\circle*{.5}}
  \put(0,19){\circle*{.5}}
  \put(32,19){\circle*{.5}}
  \put(16,31){\circle*{.5}}

\put(5,-2){\makebox(0,0){$P_1$}}
\put(27,-2){\makebox(0,0){$P_2$}}
\put(33,21){\makebox(0,0){$P_3$}}
\put(16,33){\makebox(0,0){$P_4$}}
\put(-1,21){\makebox(0,0){$P_5$}}
\put(16,20.5){\makebox(0,0){$\alpha_1+\alpha_2$}}
\put(23,16){\makebox(0,0){$-\alpha_2$}}
\put(9,16){\makebox(0,0){$-\alpha_1$}}
\put(12,8.7){\makebox(0,0){$\alpha_1$}}
\put(20,8.7){\makebox(0,0){$\alpha_2$}}

\end{picture}
\end{center}
\caption{Labelling of diagonals in type $A_2$}
\label{fig:a2}
\end{figure}

The cardinality of the set $\Phi_{\geq -1}$ is equal to
$$\ell(w_\circ)+n=\frac{n(n+1)}{2} + n = \frac{n(n+3)}{2} \ .$$
This number is also the cardinality of the set of diagonals in a
convex $(n+3)$-gon.
We shall identify these two sets as follows.
(Refer to Figure~\ref{fig:a2}.)

\begin{figure}[ht]
\begin{center}
\setlength{\unitlength}{3pt}
\begin{picture}(60,66)(0,-2)
\thicklines
  \multiput(0,20)(60,0){2}{\line(0,1){20}}
  \multiput(20,0)(0,60){2}{\line(1,0){20}}
  \multiput(0,40)(40,-40){2}{\line(1,1){20}}
  \multiput(20,0)(40,40){2}{\line(-1,1){20}}

  \multiput(20,0)(20,0){2}{\circle*{1}}
  \multiput(20,60)(20,0){2}{\circle*{1}}
  \multiput(0,20)(0,20){2}{\circle*{1}}
  \multiput(60,20)(0,20){2}{\circle*{1}}

\thinlines
\put(0,20){\line(1,0){60}}
\put(0,40){\line(1,0){60}}
\put(0,20){\line(2,-1){40}}
\put(0,40){\line(3,-1){60}}
\put(20,60){\line(2,-1){40}}

\put(30,8){\makebox(0,0){$-\alpha_1$}}
\put(30,22){\makebox(0,0){$-\alpha_2$}}
\put(30,32){\makebox(0,0){$-\alpha_3$}}
\put(30,42){\makebox(0,0){$-\alpha_4$}}
\put(30,52){\makebox(0,0){$-\alpha_5$}}

\put(40,-3){\makebox(0,0){$P_1$}}
\put(63,20){\makebox(0,0){$P_2$}}
\put(63,40){\makebox(0,0){$P_3$}}
\put(40,63){\makebox(0,0){$P_4$}}
\put(20,63){\makebox(0,0){$P_5$}}
\put(-3,40){\makebox(0,0){$P_6$}}
\put(-3,20){\makebox(0,0){$P_7$}}
\put(20,-3){\makebox(0,0){$P_8$}}

\end{picture}
\end{center}
\caption{The ``snake'' in type $A_5$}
\label{fig:octagon}
\end{figure}
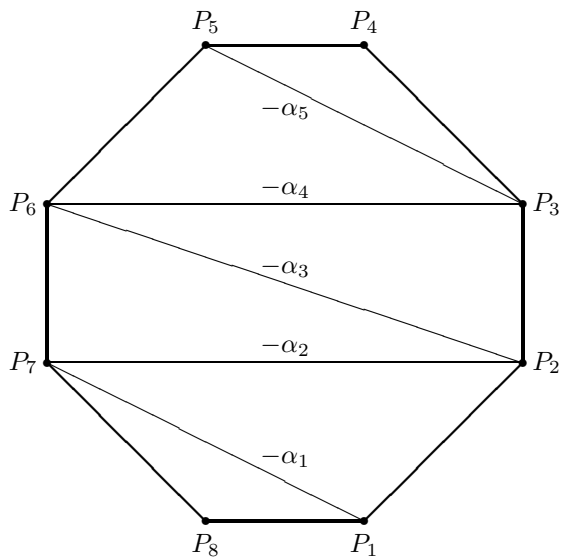

Let $P_1, \dots, P_{n+3}$ be the vertices of a regular convex
$(n+3)$-gon,
labeled counterclockwise.
For $1\leq i \leq \frac{n+1}{2}$,
we identify the root $- \alpha_{2i-1} \in \Phi_{\geq -1}$
with the diagonal $[P_i, P_{n+3-i}]$;
for $1\leq i \leq \frac{n}{2}$,
we identify the root
$- \alpha_{2i}$ with the diagonal $[P_{i+1}, P_{n+3-i}]$.
These diagonals form a ``snake" shown in Figure~\ref{fig:octagon}.
To identify the remaining diagonals (not belonging to the
snake) with positive roots, we associate each 
$\alpha_{ij}$ with the unique diagonal that crosses the diagonals
$- \alpha_i, - \alpha_{i+1}, \dots, - \alpha_j$ and does not cross
any other diagonal~$- \alpha_k$ on the snake.
(From this point on, we shall use the term ``cross'' to mean
``intersect inside the polygon.'')

Under the identification described above, all notions related to
compatible subsets and clusters of type $A_n$ can be translated
into the language of plane geometry.
The following proposition is
checked by direct inspection.


\begin{proposition}
\label{pr:clusters-A}
Let $\Phi$ be a root system of type~$A_n\,$. 
\begin{enumerate}
\item
The transformation $\tau_+$ (resp., $\tau_-$)
acts in $\Phi_{\geq -1}$ by the orthogonal  reflection of the
$(n+3)$-gon that sends each vertex $P_i$ to $P_{n+4-i}$
(resp., to $P_{n+3-i}$, with the convention $P_0 = P_{n+3}$).
Thus, $\tau_- \tau_+$ (resp., $\tau_+ \tau_-$) acts by clockwise
(resp., counter-clockwise) rotation by~$\frac{2 \pi}{n+3}$.

\item
For $\alpha, \beta \in \Phi_{\geq -1}$, the
compatibility degree $(\alpha \| \beta)$ is equal to
$1$ if the diagonals $\alpha$ and $\beta$ cross each other, and
$0$ otherwise.

\item
Compatible sets are
collections of mutually non-crossing diagonals.
Thus, clusters are in bijection with triangulations of the
$(n+3)$-gon by non-crossing diagonals
(and therefore with non-crossing partitions of $[1,n\!+\!1]$;
see, e.g., \cite[5.1]{simion}).

\item
Two triangulations are joined by an edge in the exchange graph if
and only if they are obtained from each other by a flip
which replaces a diagonal in a quadrilateral formed by two triangles
of the triangulation by another diagonal of the same quadrilateral.
\end{enumerate}
\end{proposition}

(Recall that according to Definition~\ref{def:exchange-graph},
the vertices of the exchange graph are the clusters, and two of them are
connected by an edge if they intersect by $n\!-\!1$ elements.)

The description of the exchange graph in
Proposition~\ref{pr:clusters-A} implies
Conjecture~\ref{con:cluster-polytope-weak}
for the type~$A_n\,$.
It shows that the polytope in question is the
\emph{Stasheff polytope,} or \emph{associahedron}
(see \cite{stasheff}, \cite{lee}, \cite[Chapter~7]{gkz}).

\medskip

\pagebreak[2]

\noindent {\bf Types $B_n$ and $C_n$.}
\label{sec:type-bc}
Let $\Phi$ be a root system of type $B_n\,$, and $\Phi^\vee$
the dual root system of type~$C_n$.
To describe compatible subsets for $\Phi$ and $\Phi^\vee$, we
employ the folding procedure $A_{2n-1} \to B_n$ discussed at the
beginning of Section~\ref{sec:fibonacci}.
Let $\tilde \Phi$ be a root system of type $A_{2n-1}$,
and let $\rho$ be the automorphism of $\tilde \Phi$ that sends each
simple root $\tilde \alpha_{i}$ to $\tilde \alpha_{2n-i}$.
Then each of the sets $\Phi_{\geq -1}$ and $\Phi^\vee_{\geq -1}$
can be identified with the set of $\rho$-orbits in $\tilde \Phi_{\geq -1}$.
This identification induces the labeling of simple roots of type
$B_n$ by $[1,n]$ and also the choice of a sign function
$\varepsilon$: thus, a simple root $\alpha_i$ in $\Phi$
corresponds to the $\rho$-orbit of $\tilde \alpha_i$, and
$\varepsilon (i) = (-1)^{i-1}$.

We represent the elements of $\tilde \Phi_{\geq -1}$ as diagonals
of the regular $(2n+2)$-gon, as described above in our discussion of
the type~$A$ case.
Then $\rho$ is geometrically represented by the
\emph{central symmetry} (or, equivalently, the $180^\circ$ rotation)
of the polygon, which sends each vertex $P=P_{\tilde i}$
to the antipodal vertex $-P\stackrel{\rm def}{=}P_{\langle n+1+ \tilde i \rangle}$; here
$\langle m \rangle$ denotes the element of $[1,2n+2]$ congruent to
$m$ modulo $2n+2$.
We shall refer to the diagonals that join antipodal vertices
as \emph{diameters}.
It follows that one can represent an element of
$\Phi_{\geq -1}$ (resp., $\Phi^\vee_{\geq -1}$)
either as a diameter $[P,-P]$, or as as an unordered pair of centrally
symmetric non-diameter diagonals $\{[P,Q], [-P,-Q]\}$ of the $(2n+2)$-gon.
In particular, each of the roots $- \alpha_i$
(resp., $-\alpha^\vee_i$) for $i = 1, \dots, n-1$ is identified with the pair
of diagonals representing $-\tilde \alpha_i$ and
$-  \tilde \alpha_{2n-i}$, whereas $- \alpha_n$
(resp., $-\alpha^\vee_n$) is identified with the diameter
representing~$-\tilde \alpha_n$.

The case $n=3$ of this construction is illustrated in Figure~\ref{fig:b-octagon}.

\begin{figure}[ht]
\begin{center}
\setlength{\unitlength}{3pt}
\begin{picture}(60,67)(0,-2)
\thicklines
  \multiput(0,20)(60,0){2}{\line(0,1){20}}
  \multiput(20,0)(0,60){2}{\line(1,0){20}}
  \multiput(0,40)(40,-40){2}{\line(1,1){20}}
  \multiput(20,0)(40,40){2}{\line(-1,1){20}}

  \multiput(20,0)(20,0){2}{\circle*{1}}
  \multiput(20,60)(20,0){2}{\circle*{1}}
  \multiput(0,20)(0,20){2}{\circle*{1}}
  \multiput(60,20)(0,20){2}{\circle*{1}}

\thinlines
\put(0,20){\line(1,0){60}}
\put(0,40){\line(1,0){60}}
\put(0,20){\line(2,-1){40}}
\put(0,40){\line(3,-1){60}}
\put(20,60){\line(2,-1){40}}

\put(30,8){\makebox(0,0){$-\alpha_1$}}
\put(30,22){\makebox(0,0){$-\alpha_2$}}
\put(30,32){\makebox(0,0){$-\alpha_3$}}
\put(30,42){\makebox(0,0){$-\alpha_2$}}
\put(30,52){\makebox(0,0){$-\alpha_1$}}

\put(40,-3){\makebox(0,0){$P_1$}}
\put(63,20){\makebox(0,0){$P_2$}}
\put(63,40){\makebox(0,0){$P_3$}}
\put(40,63){\makebox(0,0){$P_4$}}
\put(20,63){\makebox(0,0){$-P_1$}}
\put(-4,40){\makebox(0,0){$-P_2$}}
\put(-4,20){\makebox(0,0){$-P_3$}}
\put(20,-3){\makebox(0,0){$-P_4$}}
\end{picture}

\bigskip\bigskip\bigskip

\begin{tabular}{c|c|c}
type $B$ root & type $C$ root & diameter or pair of diagonals \\[.05in]
\hline
& & \\[-.1in]
$\alpha_1$ & $\alpha_1^\vee$ & $[\pm P_2,\mp P_4]$ \\[.1in]
$\alpha_2$ & $\alpha_2^\vee$ & $[\pm P_1,\mp P_2]$ \\[.1in]
$\alpha_1+\alpha_2$ & $\alpha_1^\vee+\alpha_2^\vee$ & $[\pm P_2,\pm P_4]$ \\[.1in]
$\alpha_2+2\alpha_3$ & $\alpha_2^\vee+\alpha_3^\vee$ & $[\pm P_1,\pm P_3]$ \\[.1in]
$\alpha_1+\alpha_2+2\alpha_3$ & $\alpha_1^\vee+\alpha_2^\vee+\alpha_3^\vee$ & $[\pm P_3,\mp P_4]$ \\[.1in]
$\alpha_1+2\alpha_2+2\alpha_3$ & $\alpha_1^\vee+2\alpha_2^\vee+\alpha_3^\vee$ & $[\pm P_1,\pm P_4]$ \\[.1in]
$\alpha_3$ & $\alpha_3^\vee$ & $[P_3,-P_3]$ \\[.1in]
$\alpha_2+\alpha_3$ & $2\alpha_2^\vee+\alpha_3^\vee$ & $[P_1,-P_1]$ \\[.1in]
$\alpha_1+\alpha_2+\alpha_3$ & $2\alpha_1^\vee+2\alpha_2^\vee+\alpha_3^\vee$ & $[P_4,-P_4]$
\end{tabular}
\end{center}
\caption{Representing the roots in $\Phi_{\geq -1}$ for the types
  $B_3$ and $C_3$}
\label{fig:b-octagon}
\end{figure}

Proposition~\ref{pr:clusters-A} implies the following.

\begin{proposition}
\label{pr:clusters-B}
Let $\Phi$ be a root system of type~$B_n\,$,
and $\Phi^\vee$ the dual root system of type~$C_n\,$. 
\begin{enumerate}
\item
The transformations $\tau_+$, $\tau_-$
and their compositions have the same geometric meaning as in
Proposition~\ref{pr:clusters-A}.1.

\item
For $\alpha, \beta \in \Phi_{\geq -1}$, the
compatibility degree
$(\alpha \| \beta) = (\beta^\vee \| \alpha^\vee)$
(cf.\ Proposition~\ref{pr:compatibility-symmetry}.1)
has the following geometric meaning:
if $[P,Q]$ is one of the diagonals representing~$\alpha$,
then $(\alpha \| \beta)$ is equal to
the number of crossings of $[P,Q]$ by the diagonals representing $\beta$.

\item
The clusters for type $B_n$ or $C_n$
are in bijection with centrally symmetric triangulations of a
regular $(2n+2)$-gon by non-crossing diagonals.

\item
Two centrally symmetric triangulations are joined by an edge in
the exchange graph $E(\Phi)$ (or $E(\Phi^\vee)$) if
and only if they are obtained from each other either by a flip
involving two diameters (see Proposition~\ref{pr:clusters-A}.4),
or by a pair of centrally symmetric flips.
\end{enumerate}
\end{proposition}

The above description of the exchange graph of type~$B$ 
shows that the corresponding simplicial complex $\Delta(\Phi)$
is identical to Simion's type~$B$ associahedron 
(see \cite[Section~5.2]{simion} and \cite{simion-B}).
The centrally symmetric triangulations that label its vertices (and
our clusters) are in bijection with noncrossing partitions of type~$B$
defined by V.~Reiner~\cite{reiner}.
As shown by S.~Devadoss~\cite{devadoss}, Simion's construction is
combinatorially equivalent 
to the ``cyclohedron'' complex of R.~Bott and C.~Taubes~\cite{bott-taubes}.
This implies Conjecture~\ref{con:cluster-polytope-weak}
for the types~$B_n$ and~$C_n$,
since the cyclohedron can be realized as a convex polytope
(see M.~Markl~\cite{markl} or R.~Simion~\cite{simion-B}).

The number of centrally symmetric triangulations of a regular
$(2n+2)$-gon by non-crossing diagonals is equal
to $(n+1) c_n=\binom{2n}{n}$, in agreement with
the type~$B_n$ entry in Table~\ref{tab:cluster-numbers}.
Indeed, any such triangulation involves
precisely one diameter; 
we have $n+1$ choices for it,  and $c_n$ ways to complete
a triangulation thereafter.

\medskip

\pagebreak[4]

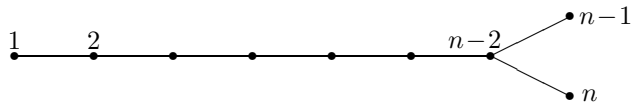
\begin{figure}[ht]

\setlength{\unitlength}{1.5pt}
\begin{picture}(180,19)(7,-7)
\put(20,0){\line(1,0){120}}
\put(140,0){\line(2,-1){20}}
\put(140,0){\line(2,1){20}}
\multiput(20,0)(20,0){7}{\circle*{2}}
\put(160,10){\circle*{2}}
\put(160,-10){\circle*{2}}
\put(20,4){\makebox(0,0){$1$}}
\put(40,4){\makebox(0,0){$2$}}
\put(136,4){\makebox(0,0){$n\!-\!2$}}
\put(169,10){\makebox(0,0){$n\!-\!1$}}
\put(165,-10){\makebox(0,0){$n$}}
\end{picture}

\caption{Coxeter graph of type $D_n$}
\label{fig:dynkin-diagram-Dn}
\end{figure}

\noindent {\bf Type $D_n$.}
Let $\Phi$ be a root system of type~$D_n\,$.
We use the following numbering of simple roots:
the indices $1, \dots, n-2$ form a chain in the Coxeter graph $I$,
while both $n-1$ and $n$ are adjacent to $n-2$.
(See Figure~\ref{fig:dynkin-diagram-Dn}.)
We also use the following sign function:
$\varepsilon (i) = (-1)^{i-1}$ for $i \in [1,n-1]$,
and $\varepsilon (n) = (-1)^{n}$.

\pagebreak[2]

The folding $D_n \to C_{n-1}$ motivates our choice of a geometric
realization of~$\Phi_{\geq -1}\,$.
Thus, we have the canonical surjection
$\pi: \Phi_{\geq -1} \to \Phi'_{\geq -1}$, where $\Phi'$ is the root
system of type~$B_{n-1}\,$.
Let $\alpha_1, \dots, \alpha_n$ be the simple roots of $\Phi$, and
$\alpha'_1, \dots, \alpha'_{n-1}$ the simple roots of $\Phi'$.
The projection $\pi$ is two-to-one over the $n$-element subset
\[
D = \{- \alpha'_{n-1}\,,\ \textstyle\sum_{j = i}^{n-1} \alpha'_j \ (1 \leq i \leq n-1)\}
\subset \Phi'_{\geq -1} \,,
\]
and one-to-one over its complement $\Phi'_{\geq -1} - D$.

As before, we identify $\Phi'_{\geq -1}$ with the set of diameters
$[P, -P]$
and unordered pairs of centrally symmetric non-diameter diagonals
$\{[P,Q], [-P,-Q]\}$ in a regular $2n$-gon.
Under this identification, $D$ becomes the set of diameters.
Motivated by the above description of the projection $\pi: \Phi_{\geq -1} \to \Phi'_{\geq -1}$,
we shall identify $\Phi_{\geq -1}$ with the disjoint union
$\Phi'_{\geq -1} \cup D'$, where $D'$ is an additional set of $n$
elements $[P,-P]'$ associated with the diameters of the $2n$-gon.
It is convenient to think that each diameter in $\Phi_{\geq -1}$ is
colored in one of two different colors.

To specify the bijection between $\Phi_{\geq -1}$ and $\Phi'_{\geq -1} \cup D'$,
we have to make a choice: for every $\alpha' \in D$,
we must decide which of the two roots $\alpha$ in $\pi^{-1} (\alpha')$
is to be identified with the diameter $[P,-P]$ representing~$\alpha'$;
the remaining element of $\pi^{-1} (\alpha')$ will then be
identified with $[P,-P]' \in D'$.
Our choice is the following: take $\alpha = - \alpha_n$ for $\alpha' = - \alpha'_{n-1}$,
and $\alpha = \alpha_i + \cdots + \alpha_{n-1}$ for
$\alpha' = \alpha'_i + \cdots + \alpha'_{n-1}$.

The case $n=4$ of this construction is illustrated in Figure~\ref{fig:d-octagon}.

\begin{figure}[ht]
\begin{center}
\setlength{\unitlength}{3pt}
\begin{picture}(60,65)(0,-2)
\thicklines
  \multiput(0,20)(60,0){2}{\line(0,1){20}}
  \multiput(20,0)(0,60){2}{\line(1,0){20}}
  \multiput(0,40)(40,-40){2}{\line(1,1){20}}
  \multiput(20,0)(40,40){2}{\line(-1,1){20}}

  \multiput(20,0)(20,0){2}{\circle*{1}}
  \multiput(20,60)(20,0){2}{\circle*{1}}
  \multiput(0,20)(0,20){2}{\circle*{1}}
  \multiput(60,20)(0,20){2}{\circle*{1}}

\thinlines
\put(0,20){\line(1,0){60}}
\put(0,40){\line(1,0){60}}
\put(0,20){\line(2,-1){40}}
\put(1,40){\line(3,-1){59}}
\put(0,39.5){\line(3,-1){4}}
\put(6,37.5){\line(3,-1){4}}
\put(12,35.5){\line(3,-1){4}}
\put(18,33.5){\line(3,-1){4}}
\put(24,31.5){\line(3,-1){4}}
\put(30,29.5){\line(3,-1){4}}
\put(36,27.5){\line(3,-1){4}}
\put(42,25.5){\line(3,-1){4}}
\put(48,23.5){\line(3,-1){4}}
\put(54,21.5){\line(3,-1){4}}
\put(20,60){\line(2,-1){40}}

\put(30,8){\makebox(0,0){$-\alpha_1$}}
\put(30,22){\makebox(0,0){$-\alpha_2$}}
\put(30,32.5){\makebox(0,0){$-\alpha_4$}}
\put(30,27.5){\makebox(0,0){$-\alpha_3$}}
\put(30,42){\makebox(0,0){$-\alpha_2$}}
\put(30,52){\makebox(0,0){$-\alpha_1$}}

\put(40,-3){\makebox(0,0){$P_1$}}
\put(63,20){\makebox(0,0){$P_2$}}
\put(63,40){\makebox(0,0){$P_3$}}
\put(40,63){\makebox(0,0){$P_4$}}
\put(20,63){\makebox(0,0){$-P_1$}}
\put(-4,40){\makebox(0,0){$-P_2$}}
\put(-4,20){\makebox(0,0){$-P_3$}}
\put(20,-3){\makebox(0,0){$-P_4$}}
\end{picture}

\bigskip\bigskip\bigskip

\begin{tabular}{c|c|c}
type $B$ root & type $D$ root & diameter or pair of diagonals \\[.05in]
\hline
& & \\[-.1in]
$-\alpha_1$ & $-\alpha_1$ & $[\pm P_1,\mp P_3]$ \\[.1in]
$-\alpha_2$ & $-\alpha_2$ & $[\pm P_2,\mp P_3]$ \\[.1in]
$\alpha_1$ & $\alpha_1$ & $[\pm P_2,\mp P_4]$ \\[.1in]
$\alpha_2$ & $\alpha_2$ & $[\pm P_1,\mp P_2]$ \\[.1in]
$\alpha_1+\alpha_2$ & $\alpha_1+\alpha_2$ & $[\pm P_2,\pm P_4]$ \\[.1in]
$\alpha_2+2\alpha_3$ & $\alpha_2+\alpha_3+\alpha_4$ & $[\pm P_1,\pm P_3]$ \\[.1in]
$\alpha_1+\alpha_2+2\alpha_3$ & $\alpha_1+\alpha_2+\alpha_3+\alpha_4$ & $[\pm P_3,\mp P_4]$ \\[.1in]
$\alpha_1+2\alpha_2+2\alpha_3$ & $\alpha_1+2\alpha_2+\alpha_3+\alpha_4$ & $[\pm P_1,\pm P_4]$ \\[.1in]
$-\alpha_3$ & $-\alpha_4$ & $[P_2,-P_2]$ \\[.1in]
            & $-\alpha_3$ & $[P_2,-P_2]'$ \\[.1in]
$\alpha_3$ & $\alpha_3$ & $[P_3,-P_3]$ \\[.1in]
           & $\alpha_4$ & $[P_3,-P_3]'$ \\[.1in]
$\alpha_2+\alpha_3$ & $\alpha_2+\alpha_3$ & $[P_1,-P_1]$ \\[.1in]
                    & $\alpha_2+\alpha_4$ & $[P_1,-P_1]'$ \\[.1in]
$\alpha_1+\alpha_2+\alpha_3$ & $\alpha_1+\alpha_2+\alpha_3$ & $[P_4,-P_4]$ \\[.1in]
                             & $\alpha_1+\alpha_2+\alpha_4$ & $[P_4,-P_4]$
\end{tabular}
\end{center}
\caption{Representing the roots in $\Phi_{\geq -1}$ for the type $D_4$}
\label{fig:d-octagon}
\end{figure}

The following type~$D_n$ counterpart of
Propositions~\ref{pr:clusters-A}--\ref{pr:clusters-B} is
verified by a direct inspection.

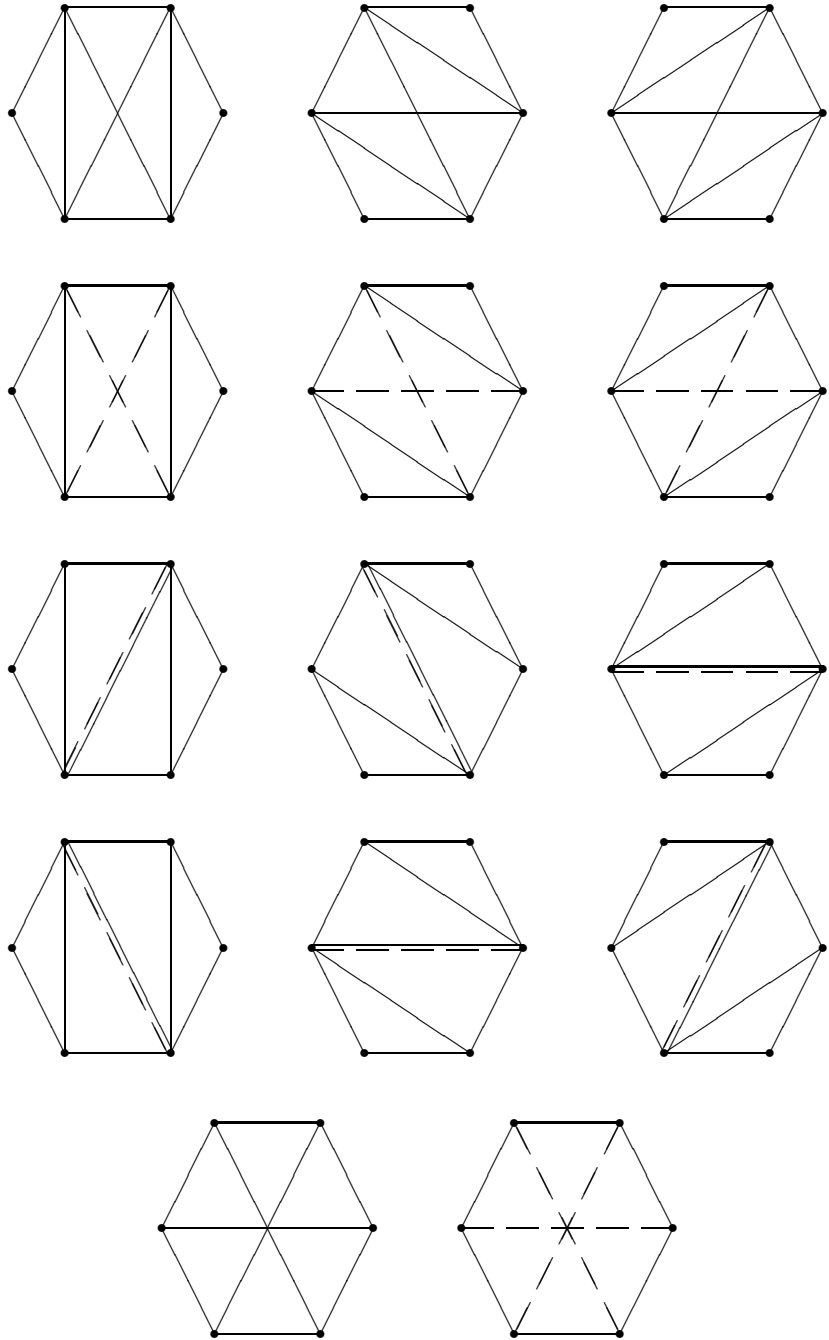
\begin{figure}[ht]
\begin{center}
\vspace{.2in}

\begin{tabular}{ccc}
\setlength{\unitlength}{2pt}
\begin{picture}(50,40)(-5,0)

  \put(0,20){\circle*{1.5}}
  \put(40,20){\circle*{1.5}}
  \put(10,0){\circle*{1.5}}
  \put(10,40){\circle*{1.5}}
  \put(30,0){\circle*{1.5}}
  \put(30,40){\circle*{1.5}}

\put(0,20){\line(1,2){10}}
\put(0,20){\line(1,-2){10}}
\put(40,20){\line(-1,2){10}}
\put(40,20){\line(-1,-2){10}}
\put(10,0){\line(1,0){20}}
\put(10,40){\line(1,0){20}}

\put(10,0){\line(1,2){20}}
\put(10,40){\line(1,-2){20}}


\put(10,0){\line(0,1){40}} \put(30,0){\line(0,1){40}}

\end{picture}
&
\setlength{\unitlength}{2pt}
\begin{picture}(50,40)(-5,0)

  \put(0,20){\circle*{1.5}}
  \put(40,20){\circle*{1.5}}
  \put(10,0){\circle*{1.5}}
  \put(10,40){\circle*{1.5}}
  \put(30,0){\circle*{1.5}}
  \put(30,40){\circle*{1.5}}

\put(0,20){\line(1,2){10}}
\put(0,20){\line(1,-2){10}}
\put(40,20){\line(-1,2){10}}
\put(40,20){\line(-1,-2){10}}
\put(10,0){\line(1,0){20}}
\put(10,40){\line(1,0){20}}

\put(0,20){\line(1,0){40}}
\put(10,40){\line(1,-2){20}}


\put(0,20){\line(3,-2){30}} \put(10,40){\line(3,-2){30}}

\end{picture}

&
\setlength{\unitlength}{2pt}
\begin{picture}(50,40)(-5,0)

  \put(0,20){\circle*{1.5}}
  \put(40,20){\circle*{1.5}}
  \put(10,0){\circle*{1.5}}
  \put(10,40){\circle*{1.5}}
  \put(30,0){\circle*{1.5}}
  \put(30,40){\circle*{1.5}}

\put(0,20){\line(1,2){10}}
\put(0,20){\line(1,-2){10}}
\put(40,20){\line(-1,2){10}}
\put(40,20){\line(-1,-2){10}}
\put(10,0){\line(1,0){20}}
\put(10,40){\line(1,0){20}}

\put(0,20){\line(1,0){40}}
\put(10,0){\line(1,2){20}}


\put(0,20){\line(3,2){30}} \put(10,0){\line(3,2){30}}

\end{picture}

\\[.3in]

\setlength{\unitlength}{2pt}
\begin{picture}(50,40)(-5,0)

  \put(0,20){\circle*{1.5}}
  \put(40,20){\circle*{1.5}}
  \put(10,0){\circle*{1.5}}
  \put(10,40){\circle*{1.5}}
  \put(30,0){\circle*{1.5}}
  \put(30,40){\circle*{1.5}}

\put(0,20){\line(1,2){10}}
\put(0,20){\line(1,-2){10}}
\put(40,20){\line(-1,2){10}}
\put(40,20){\line(-1,-2){10}}
\put(10,0){\line(1,0){20}}
\put(10,40){\line(1,0){20}}

%
\multiput(10,0)(4.25,8.5){5}{\line(1,2){3}}
\multiput(10,40)(4.25,-8.5){5}{\line(1,-2){3}}


\put(10,0){\line(0,1){40}} \put(30,0){\line(0,1){40}}

\end{picture}
&
\setlength{\unitlength}{2pt}
\begin{picture}(50,40)(-5,0)

  \put(0,20){\circle*{1.5}}
  \put(40,20){\circle*{1.5}}
  \put(10,0){\circle*{1.5}}
  \put(10,40){\circle*{1.5}}
  \put(30,0){\circle*{1.5}}
  \put(30,40){\circle*{1.5}}

\put(0,20){\line(1,2){10}}
\put(0,20){\line(1,-2){10}}
\put(40,20){\line(-1,2){10}}
\put(40,20){\line(-1,-2){10}}
\put(10,0){\line(1,0){20}}
\put(10,40){\line(1,0){20}}

\multiput(0,20)(8.5,0){5}{\line(1,0){6}}
\multiput(10,40)(4.25,-8.5){5}{\line(1,-2){3}}


\put(0,20){\line(3,-2){30}} \put(10,40){\line(3,-2){30}}

\end{picture}

&
\setlength{\unitlength}{2pt}
\begin{picture}(50,40)(-5,0)

  \put(0,20){\circle*{1.5}}
  \put(40,20){\circle*{1.5}}
  \put(10,0){\circle*{1.5}}
  \put(10,40){\circle*{1.5}}
  \put(30,0){\circle*{1.5}}
  \put(30,40){\circle*{1.5}}

\put(0,20){\line(1,2){10}}
\put(0,20){\line(1,-2){10}}
\put(40,20){\line(-1,2){10}}
\put(40,20){\line(-1,-2){10}}
\put(10,0){\line(1,0){20}}
\put(10,40){\line(1,0){20}}

\multiput(0,20)(8.5,0){5}{\line(1,0){6}}
\multiput(10,0)(4.25,8.5){5}{\line(1,2){3}}


\put(0,20){\line(3,2){30}} \put(10,0){\line(3,2){30}}

\end{picture}

\\[.3in]

\setlength{\unitlength}{2pt}
\begin{picture}(50,40)(-5,0)

  \put(0,20){\circle*{1.5}}
  \put(40,20){\circle*{1.5}}
  \put(10,0){\circle*{1.5}}
  \put(10,40){\circle*{1.5}}
  \put(30,0){\circle*{1.5}}
  \put(30,40){\circle*{1.5}}

\put(0,20){\line(1,2){10}}
\put(0,20){\line(1,-2){10}}
\put(40,20){\line(-1,2){10}}
\put(40,20){\line(-1,-2){10}}
\put(10,0){\line(1,0){20}}
\put(10,40){\line(1,0){20}}

%
\put(10.5,-0.25){\line(1,2){20}}
\multiput(9.5,0.25)(4.25,8.5){5}{\line(1,2){3}}


\put(10,0){\line(0,1){40}} \put(30,0){\line(0,1){40}}

\end{picture}
&
\setlength{\unitlength}{2pt}
\begin{picture}(50,40)(-5,0)

  \put(0,20){\circle*{1.5}}
  \put(40,20){\circle*{1.5}}
  \put(10,0){\circle*{1.5}}
  \put(10,40){\circle*{1.5}}
  \put(30,0){\circle*{1.5}}
  \put(30,40){\circle*{1.5}}

\put(0,20){\line(1,2){10}}
\put(0,20){\line(1,-2){10}}
\put(40,20){\line(-1,2){10}}
\put(40,20){\line(-1,-2){10}}
\put(10,0){\line(1,0){20}}
\put(10,40){\line(1,0){20}}

\put(10.5,40.25){\line(1,-2){20}}
\multiput(9.5,39.75)(4.25,-8.5){5}{\line(1,-2){3}}


\put(0,20){\line(3,-2){30}} \put(10,40){\line(3,-2){30}}

\end{picture}

&
\setlength{\unitlength}{2pt}
\begin{picture}(50,40)(-5,0)

  \put(0,20){\circle*{1.5}}
  \put(40,20){\circle*{1.5}}
  \put(10,0){\circle*{1.5}}
  \put(10,40){\circle*{1.5}}
  \put(30,0){\circle*{1.5}}
  \put(30,40){\circle*{1.5}}

\put(0,20){\line(1,2){10}}
\put(0,20){\line(1,-2){10}}
\put(40,20){\line(-1,2){10}}
\put(40,20){\line(-1,-2){10}}
\put(10,0){\line(1,0){20}}
\put(10,40){\line(1,0){20}}

\put(0,20.5){\line(1,0){40}}
\multiput(0,19.5)(8.5,0){5}{\line(1,0){6}}


\put(0,20){\line(3,2){30}} \put(10,0){\line(3,2){30}}

\end{picture}

\\[.3in]

\setlength{\unitlength}{2pt}
\begin{picture}(50,40)(-5,0)

  \put(0,20){\circle*{1.5}}
  \put(40,20){\circle*{1.5}}
  \put(10,0){\circle*{1.5}}
  \put(10,40){\circle*{1.5}}
  \put(30,0){\circle*{1.5}}
  \put(30,40){\circle*{1.5}}

\put(0,20){\line(1,2){10}}
\put(0,20){\line(1,-2){10}}
\put(40,20){\line(-1,2){10}}
\put(40,20){\line(-1,-2){10}}
\put(10,0){\line(1,0){20}}
\put(10,40){\line(1,0){20}}

\put(10.5,40.25){\line(1,-2){20}}
\multiput(9.5,39.75)(4.25,-8.5){5}{\line(1,-2){3}}


\put(10,0){\line(0,1){40}} \put(30,0){\line(0,1){40}}

\end{picture}
&
\setlength{\unitlength}{2pt}
\begin{picture}(50,40)(-5,0)

  \put(0,20){\circle*{1.5}}
  \put(40,20){\circle*{1.5}}
  \put(10,0){\circle*{1.5}}
  \put(10,40){\circle*{1.5}}
  \put(30,0){\circle*{1.5}}
  \put(30,40){\circle*{1.5}}

\put(0,20){\line(1,2){10}}
\put(0,20){\line(1,-2){10}}
\put(40,20){\line(-1,2){10}}
\put(40,20){\line(-1,-2){10}}
\put(10,0){\line(1,0){20}}
\put(10,40){\line(1,0){20}}

\put(0,20.5){\line(1,0){40}}
\multiput(0,19.5)(8.5,0){5}{\line(1,0){6}}


\put(0,20){\line(3,-2){30}} \put(10,40){\line(3,-2){30}}

\end{picture}

&
\setlength{\unitlength}{2pt}
\begin{picture}(50,40)(-5,0)

  \put(0,20){\circle*{1.5}}
  \put(40,20){\circle*{1.5}}
  \put(10,0){\circle*{1.5}}
  \put(10,40){\circle*{1.5}}
  \put(30,0){\circle*{1.5}}
  \put(30,40){\circle*{1.5}}

\put(0,20){\line(1,2){10}}
\put(0,20){\line(1,-2){10}}
\put(40,20){\line(-1,2){10}}
\put(40,20){\line(-1,-2){10}}
\put(10,0){\line(1,0){20}}
\put(10,40){\line(1,0){20}}

\put(10.5,-0.25){\line(1,2){20}}
\multiput(9.5,0.25)(4.25,8.5){5}{\line(1,2){3}}


\put(0,20){\line(3,2){30}} \put(10,0){\line(3,2){30}}

\end{picture}

\end{tabular}

\vspace{.3in}

\begin{tabular}{cc}

\setlength{\unitlength}{2pt}
\begin{picture}(50,40)(-5,0)

  \put(0,20){\circle*{1.5}}
  \put(40,20){\circle*{1.5}}
  \put(10,0){\circle*{1.5}}
  \put(10,40){\circle*{1.5}}
  \put(30,0){\circle*{1.5}}
  \put(30,40){\circle*{1.5}}

\put(0,20){\line(1,2){10}}
\put(0,20){\line(1,-2){10}}
\put(40,20){\line(-1,2){10}}
\put(40,20){\line(-1,-2){10}}
\put(10,0){\line(1,0){20}}
\put(10,40){\line(1,0){20}}

\put(0,20){\line(1,0){40}}
\put(10,0){\line(1,2){20}}
\put(10,40){\line(1,-2){20}}



\end{picture}
&
\setlength{\unitlength}{2pt}
\begin{picture}(50,40)(-5,0)

  \put(0,20){\circle*{1.5}}
  \put(40,20){\circle*{1.5}}
  \put(10,0){\circle*{1.5}}
  \put(10,40){\circle*{1.5}}
  \put(30,0){\circle*{1.5}}
  \put(30,40){\circle*{1.5}}

\put(0,20){\line(1,2){10}}
\put(0,20){\line(1,-2){10}}
\put(40,20){\line(-1,2){10}}
\put(40,20){\line(-1,-2){10}}
\put(10,0){\line(1,0){20}}
\put(10,40){\line(1,0){20}}

\multiput(0,20)(8.5,0){5}{\line(1,0){6}}
\multiput(10,0)(4.25,8.5){5}{\line(1,2){3}}
\multiput(10,40)(4.25,-8.5){5}{\line(1,-2){3}}



\end{picture}

\end{tabular}

\vspace{.1in}

\end{center}
\caption{Triangulations representing the clusters of type $D_3$}
\label{fig:d-hexagon}
\end{figure}

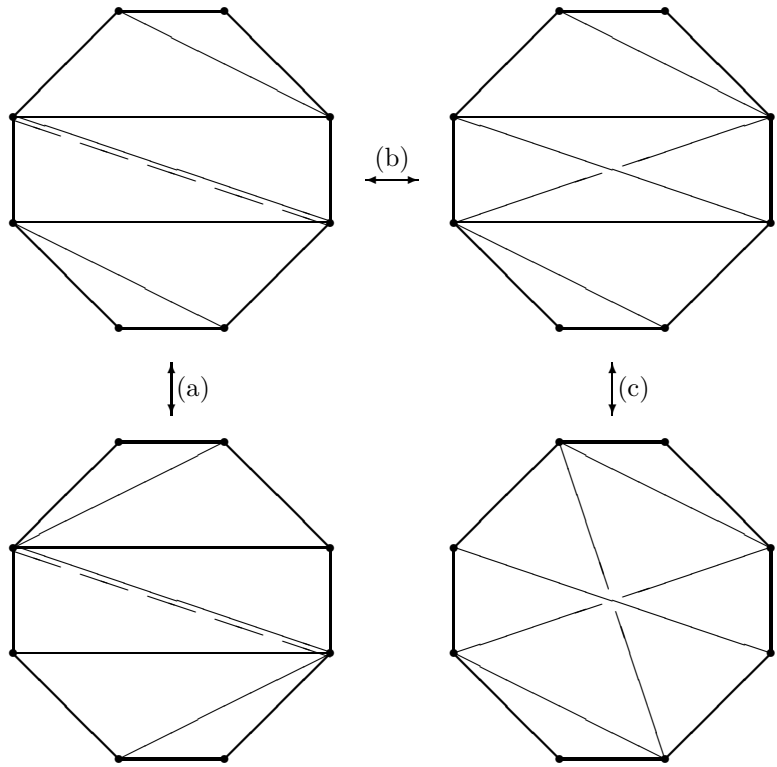
\begin{figure}[ht]
\begin{center}
\setlength{\unitlength}{2pt}
\begin{picture}(60,65)(0,-2)
\thicklines
  \multiput(0,20)(60,0){2}{\line(0,1){20}}
  \multiput(20,0)(0,60){2}{\line(1,0){20}}
  \multiput(0,40)(40,-40){2}{\line(1,1){20}}
  \multiput(20,0)(40,40){2}{\line(-1,1){20}}

  \multiput(20,0)(20,0){2}{\circle*{1.5}}
  \multiput(20,60)(20,0){2}{\circle*{1.5}}
  \multiput(0,20)(0,20){2}{\circle*{1.5}}
  \multiput(60,20)(0,20){2}{\circle*{1.5}}

\thinlines
\put(0,20){\line(1,0){60}}
\put(0,40){\line(1,0){60}}
\put(0,20){\line(2,-1){40}}
\put(1,40){\line(3,-1){59}}
\put(6,37.5){\line(3,-1){4}}
\put(12,35.5){\line(3,-1){4}}
\put(18,33.5){\line(3,-1){4}}
\put(24,31.5){\line(3,-1){4}}
\put(30,29.5){\line(3,-1){4}}
\put(36,27.5){\line(3,-1){4}}
\put(42,25.5){\line(3,-1){4}}
\put(48,23.5){\line(3,-1){4}}
\put(54,21.5){\line(3,-1){4}}
\multiput(0,39.3)(12.75,-4.25){5}{\line(3,-1){9}}
\put(20,60){\line(2,-1){40}}

\end{picture}
\begin{picture}(20,60)(0,0)
\put(10,30){\vector(1,0){5}}
\put(10,30){\vector(-1,0){5}}
\put(10,34){\makebox(0,0){(b)}}

\end{picture}
\begin{picture}(60,65)(0,-2)
\thicklines
  \multiput(0,20)(60,0){2}{\line(0,1){20}}
  \multiput(20,0)(0,60){2}{\line(1,0){20}}
  \multiput(0,40)(40,-40){2}{\line(1,1){20}}
  \multiput(20,0)(40,40){2}{\line(-1,1){20}}

  \multiput(20,0)(20,0){2}{\circle*{1.5}}
  \multiput(20,60)(20,0){2}{\circle*{1.5}}
  \multiput(0,20)(0,20){2}{\circle*{1.5}}
  \multiput(60,20)(0,20){2}{\circle*{1.5}}

\thinlines
\put(0,20){\line(1,0){60}}
\put(0,40){\line(1,0){60}}
\put(0,20){\line(2,-1){40}}
\put(0,40){\line(3,-1){60}}
\put(0,20){\line(3,1){28}}
\put(60,40){\line(-3,-1){28}}
\put(20,60){\line(2,-1){40}}

\end{picture}
\end{center}


\begin{center}
\setlength{\unitlength}{2pt}
\begin{picture}(60,80)(0,-2)
\thicklines
  \multiput(0,20)(60,0){2}{\line(0,1){20}}
  \multiput(20,0)(0,60){2}{\line(1,0){20}}
  \multiput(0,40)(40,-40){2}{\line(1,1){20}}
  \multiput(20,0)(40,40){2}{\line(-1,1){20}}

  \multiput(20,0)(20,0){2}{\circle*{1.5}}
  \multiput(20,60)(20,0){2}{\circle*{1.5}}
  \multiput(0,20)(0,20){2}{\circle*{1.5}}
  \multiput(60,20)(0,20){2}{\circle*{1.5}}

\thinlines
\put(0,20){\line(1,0){60}}
\put(0,40){\line(1,0){60}}
\put(20,0){\line(2,1){40}}
\put(1,40){\line(3,-1){59}}
\put(0,39.5){\line(3,-1){4}}
\put(6,37.5){\line(3,-1){4}}
\put(12,35.5){\line(3,-1){4}}
\put(18,33.5){\line(3,-1){4}}
\put(24,31.5){\line(3,-1){4}}
\put(30,29.5){\line(3,-1){4}}
\put(36,27.5){\line(3,-1){4}}
\put(42,25.5){\line(3,-1){4}}
\put(48,23.5){\line(3,-1){4}}
\put(54,21.5){\line(3,-1){4}}
\multiput(0,39.3)(12.75,-4.25){5}{\line(3,-1){9}}
\put(0,40){\line(2,1){40}}

\put(30,70){\vector(0,1){5}}
\put(30,70){\vector(0,-1){5}}
\put(34,70){\makebox(0,0){(a)}}

\end{picture}
\begin{picture}(20,60)(0,0)

\end{picture}
\begin{picture}(60,80)(0,-2)
\thicklines
  \multiput(0,20)(60,0){2}{\line(0,1){20}}
  \multiput(20,0)(0,60){2}{\line(1,0){20}}
  \multiput(0,40)(40,-40){2}{\line(1,1){20}}
  \multiput(20,0)(40,40){2}{\line(-1,1){20}}

  \multiput(20,0)(20,0){2}{\circle*{1.5}}
  \multiput(20,60)(20,0){2}{\circle*{1.5}}
  \multiput(0,20)(0,20){2}{\circle*{1.5}}
  \multiput(60,20)(0,20){2}{\circle*{1.5}}

\thinlines
\put(0,20){\line(2,-1){40}}
\put(0,40){\line(3,-1){60}}
\put(0,20){\line(3,1){28}}
\put(60,40){\line(-3,-1){28}}
\put(20,60){\line(1,-3){9.3}}
\put(40,0){\line(-1,3){9.3}}

\put(20,60){\line(2,-1){40}}

\put(30,70){\vector(0,1){5}}
\put(30,70){\vector(0,-1){5}}
\put(34,70){\makebox(0,0){(c)}}

\end{picture}
\end{center}

\caption{Flips of type $D$}
\label{fig:d-flips}
\end{figure}

\begin{proposition}
\label{pr:clusters-D}
Let $\Phi$ be a root system of type~$D_n\,$.
\begin{enumerate}
\item
The transformations $\tau_+$ and $\tau_-$ of
$\Phi_{\geq -1}$ are realized by the same reflections as in
Propositions~\ref{pr:clusters-A}.1
and~\ref{pr:clusters-B}.1, with the following
modification: 
$\tau_{(-1)^n}$ involves changing the colors of all diameters.

\item
For $\alpha, \beta \in \Phi_{\geq -1}$, the
compatibility degree
$(\alpha \| \beta) = (\beta \| \alpha)$
has the following geometric meaning:
\begin{itemize}
\item
if 
$\alpha\notin D \cup D'$, $\beta\notin D \cup D'$, 
then
$(\alpha \| \beta)$ is the same as in
Proposition~\ref{pr:clusters-B}.2;

\item
if, say, $\alpha \!\in\! D \cup D'$, $\beta\! \notin\! D \cup D'$, then
$(\alpha \| \beta)\! =\! 1$ if the diameter representing $\alpha$
crosses both diagonals representing $\beta$; otherwise,
$(\alpha \| \beta)\! =\! 0$;

\item
if $\alpha, \beta \in D$ or
$\alpha, \beta \in D'$, then $(\alpha \| \beta) = 0$;

\item
if $\alpha \in D$ and $\beta \in D'$, then
$(\alpha \| \beta) = 1$ if the diameters representing $\alpha$ and
$\beta$ are different; otherwise, $(\alpha \| \beta) = 0$.
\end{itemize}
\item
The clusters for type $D_n$
are in bijection with centrally symmetric colored triangulations of the
regular $2n$-gon which fit the following description
(see Figure~\ref{fig:d-hexagon}):
\begin{itemize}
\item each triangulation is made of 
  non-diameter diagonals together with at least two colored diameters
  (that is, elements of either $D$ or~$D'$);
\item 
the diagonals making up a
  triangulation do not have common internal points, except for diameters of the same color
  crossing at the center, or diameters of different color connecting
  the same antipodal points.
\end{itemize}


\item
Two triangulations of the kind described in Part~3 above are joined by
an edge in the exchange graph
$E(\Phi)$ if and only if they are obtained from each other
by one of the ``type~$D$ flips'' shown in
Figure~\ref{fig:d-flips}:
\begin{itemize}
\item[{\rm (a)}]
a pair of centrally symmetric flips
(cf.\ Proposition~\ref{pr:clusters-A}.4);
\item[{\rm (b)}]
a flip involving two diameters of different colors;
\item[{\rm (c)}]
a ``hexagonal
flip" exchanging a diameter 
with a pair of diagonals.
\end{itemize}
\end{enumerate}
\end{proposition}

\begin{thebibliography}{xxx}

\bibitem{ath}
C.~Athanasiadis,
On noncrossing and nonnesting partitions for classical reflection
groups, \textsl{Electron.\ J.\ Combin.} \textbf{5} (1998), Research
Paper 42, 16 pp. (electronic).

\bibitem{bfz96}
A.~Berenstein, S.~Fomin and A.~Zelevinsky,
Parametrizations of canonical bases and totally positive matrices,
\textsl{Adv.\ Math.} {\bf 122} (1996), 49--149.

\bibitem{bz01}
A.~Berenstein and A.~Zelevinsky,
Tensor product multiplicities, canonical bases and totally positive varieties,
\textsl{Invent.\ Math.} \textbf {143} (2001), 77--128.

\bibitem{bott-taubes}
R.~Bott and C.~Taubes,
On the self-linking of knots. Topology and physics,
\textsl{J.~Math.\ Phys.} \textbf{35} (1994), no. 10, 5247--5287.

\bibitem{bourbaki} N.~Bourbaki,
{\sl Groupes et alg\`ebres de Lie},
Ch.~IV-VI, Hermann, Paris, 1968.

\bibitem{CGT}
R.~Caracciolo, F.~Gliozzi, and R.~Tateo,
A topological invariant of RG flows in $2$D integrable quantum field
theories,
\textsl{Internat.\ J.\ Modern Phys.~B} \textbf{13} (1999), no.~24-25, 2927--2932.

\bibitem{devadoss}
S.~L.~Devadoss, A space of cyclohedra, preprint
\texttt{math.QA/0102166}.

\bibitem{fz-clust1}
S.~Fomin and A.~Zelevinsky,
Cluster algebras I: Foundations,
\textsl{J.~Amer.\ Math.\ Soc.}, to appear.


\bibitem{fz-Laurent}
S.~Fomin and A.~Zelevinsky,
The Laurent phenomenon,
\textsl{Adv.\ in Appl.\ Math.}, to appear.

\bibitem{FSS}
L.~Frappat, A.~Sciarrino, and P.~Sorba,
\textsl{Dictionary on Lie algebras and superalgebras,}
Academic Press, San Diego, CA, 2000.

\bibitem{frenkel-szenes}
E.~Frenkel and A.~Szenes,
Thermodynamic Bethe ansatz and dilogarithm identities.~I,
\textsl{Math.\ Res.\ Lett.} \textbf{2} (1995), no.~6, 677--693.

\bibitem{gkz}
I.~Gelfand, M.~Kapranov, and A.~Zelevinsky,
\textsl{Discriminants, Resultants and Multidimensional Determinants,}
Birkh\"auser Boston, 1994, 523 pp.

\bibitem{gliozzi-tateo}
F.~Gliozzi and  R.~Tateo,
Thermodynamic Bethe ansatz and three-fold triangulations,
\textsl{Internat.\ J.\ Modern Phys.~A} \textbf{11} (1996), no.~22, 4051--4064.


\bibitem{kuniba-nakanishi}
A.~Kuniba and T.~Nakanishi,
Spectra in conformal field theories from the Rogers dilogarithm,
\textsl{Modern Phys.\ Lett.~A} \textbf{7} (1992), no.~37, 3487--3494.

\bibitem{KNS}
A.~Kuniba, T.~Nakanishi, and J.~Suzuki,
Functional relations in solvable lattice models.~I.
Functional relations and representation theory,
\textsl{Internat.\ J.\ Modern Phys.~A} \textbf{9} (1994), no.~30,
5215--5266.

\bibitem{lee}
C.~W.~Lee,
The associahedron and triangulations of the $n$-gon,
\textsl{European J.\ Combin.} \textbf{10} (1989), no. 6, 551--560.

\bibitem{markl}
M.~Markl,
Simplex, associahedron, and cyclohedron,
\textsl{Contemp.\ Math.} \textbf{227} (1999),
235--265.

\bibitem{RVT}
F.~Ravanini, A.~Valleriani, and R.~Tateo,
Dynkin TBAs,
\textsl{Internat.\ J.\ Modern Phys.~A} \textbf{8} (1993), no.~10, 1707--1727.

\bibitem{reiner}
V.~Reiner,
Non-crossing partitions for classical reflection groups,
\textsl{Discrete Math.} \textbf{177}  (1997), 195--222.

\bibitem{simion}
R.~Simion,
Noncrossing partitions, \textsl{Discrete Math.}
\textbf{217} (2000), no. 1-3, 367--409.

\bibitem{simion-B}
R.~Simion,
A type-B associahedron, \textsl{Adv.\ in Appl.\ Math.}, to appear.

\bibitem{stasheff}
J.~D.~Stasheff,
Homotopy associativity of $H$-spaces.~I,~II,
\textsl{Trans.\ Amer.\ Math.\ Soc.} \textbf{108} (1963), 275--292, 293--312.

\bibitem{zamolodchikov}
Al.~B.~Zamolodchikov,
On the thermodynamic Bethe ansatz equations for reflectionless $ADE$
scattering theories,
\textsl{Phys.\ Lett.~B} \textbf{253} (1991), 391--394.

\bibitem{ziegler}
G.~M.~Ziegler,
\textsl{Lectures on polytopes,}
Springer-Verlag, New York, 1995.

\end{thebibliography}
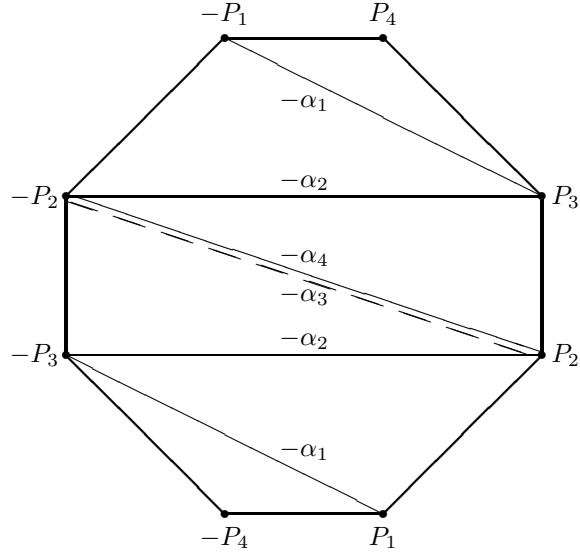
\end{document}